\documentstyle[aaspp4,11pt]{article}

\newcommand{\be}{\begin{eqnarray}}
\newcommand{\ee}{\end{eqnarray}}

\def\etal{{\it et al.}}
\def\simless{\mathbin{\lower 3pt\hbox
   {$\rlap{\raise 5pt\hbox{$\char'074$}}\mathchar"7218$}}} %< or of order
\def\simgreat{\mathbin{\lower 3pt\hbox
   {$\rlap{\raise 5pt\hbox{$\char'076$}}\mathchar"7218$}}} %> or of order

\def\cnsq{{C_n^2}} 

\def\dne{{\delta n_e}}
\def\pne{{P_{\delta n_e}}}

\def\eps02{{{\epsilon_{-2}}}}

\def\rns{{R_{\rm NS}}}
\def\kms{{km s$^{-1}$}}

\def\mura{{\mu_{\rm RA} }}
\def\mudec{{\mu_{\rm DEC} }}

\def\dhat{{\hat D}}

\def\vperp{{V_{\perp}}}
\def\vpperp{{{V_p}_{\perp}}}

\def\dnud{{\Delta\nu_{\rm d}}}
\def\dnu{{\delta\nu}}
\def\Dnu{{\Delta\nu}}
\def\dtd{{\Delta t_{\rm d}}}
\def\dt{{\delta t}}

\def\Thetavec{{ \mbox{\boldmath $\Theta$} }}
\def\Thvec{{ \mbox{\boldmath $\Thetavec$} }}
\def\thvec{{ \mbox{\boldmath $\theta$} }}

\def\thivec{{ \mbox{\boldmath $\theta_i$} }}
\def\avec{{ \mbox{\boldmath $a$} }}
\def\th{{\theta}}

\def\ths{{\theta_d}}

\def\thmax{{\theta_{\rm max}}}
\def\taumax{{\tau_{\rm max}}}
\def\taubar{{\overline{\tau}}}
\def\taui{{\tau_i}}
\def\taup{{\tau^{\prime}}}

\def\muvec{{ \mbox{\boldmath $\mu$} }}

\def\xvec{{\bf x}}
\def\yvec{{\bf y}}
\def\dxvec{{ \mbox{\boldmath $\delta x$} }}
\def\dxvece{{ \mbox{\boldmath $\delta x_e$} }}
\def\dxperpvec{{ \mbox{\boldmath $\delta x_{\perp}$} }}

\def\dxpvec{{ \mbox{\boldmath $\Delta x^{\prime}$} }}
\def\ld{{\ell_d}}

\def\wc{{W_{\rm C}}}
\def\wdiss{{W_{\rm D,ISS}}}
\def\wdpm{{W_{\rm D,PM}}}
\def\sp{{s^{\prime}}}
\def\spp{{s^{\prime\prime}}}
\def\dphi{{D_{\phi}}}
\def\signe{{\sigma_{n_e}}}
\def\wpm{{W_{\rm PM}}}

\def\vismperp{{ {V_{\rm m}}_\perp}}

\def\vp{{V_{\rm p}}}
\def\vobs{{V_{\rm obs}}}
\def\vism{{V_{\rm m}}}

\def\vpvec{{\bf V_{\rm p}}}
\def\vobsvec{{\bf V_{\rm obs}}}
\def\vismvec{{\bf V_{\rm m}}}
\def\DV{{\bf \Delta V}}

\def\vpperpvec{{\bf   {V_{\rm p}}_\perp}}
\def\vobsperpvec{{\bf {V_{\rm obs}}_\perp}}
\def\vismperpvec{{\bf {V_{\rm m}}_\perp}}

\def\nhat{{\bf \hat n}}
\def\lhat{{ \mbox{\boldmath $\hat\ell$}}}
\def\mhat{{\bf \hat m}}
\def\veffvec{{\bf V_{\rm eff}}}
\def\veffperpvec{{\bf V_{\rm eff,\perp}}}
\def\veff{{\rm V_{\rm eff}}}

\def\dvgvec{{ \mbox{\boldmath $\Delta V_g$}}}
\def\dvg{{\Delta V_g}}

\def\dvgl{{{\Delta V_g}_{\ell}}}
\def\dvgm{{{\Delta V_g}_{m}}}
\def\dvgperp{{{\Delta V_g}_{\perp}}}
\def\dvgperpvec{{{\dvgvec}_{\perp}}}
\def\asun{{a_{\odot}}}
\def\veffvecg{{{\veffvec}_{g}}} 
\def\veffperpvecg{{{\veffvec}_{g\perp}}} 

\def\dvismperpvec{{\delta\vismperpvec}}

\def\viss{{V_{\rm ISS}}}
\def\visssq{{V^2_{\rm ISS}}}
\def\vissu{{V_{\rm ISS, u}}}

\def\vissKolu{{V_{\rm ISS, 5/3, u}}}	% uniform Kolmogorov medium
\def\vissSQu{{V_{\rm ISS, 2, u}}}	% uniform sq.-law medium
\def\aiss{{A_{\rm ISS}}}
\def\aissu{{A_{\rm ISS,u}}}

\def\aissKolu{{A_{\rm ISS, 5/3, u}}}	% uniform Kolmogorov medium
\def\DissKolu{{D_{\rm ISS, 5/3, u}}}	% uniform Kolmogorov medium
\def\dpo{{d}}				% pulsar-observer distance
\def\ds{{D_s}}				% pulsar-screen distance
\def\dso{{D}}				% screen-observer distance
\def\dds{{D - D_s}}

\def\c1u{{C_{1,u}}}
\def\c1Kolu{{C_{1,5/3,u}}}

\def\half{{\frac{1}{2}}}

\def\Omegavecperp{{ \mbox{\boldmath $\Omega_{\perp}$} }}
\def\Omegavecperphat{{ \mbox{\boldmath $\hat{\Omega}_{\perp}$} }}
\def\mathbf{{\bf}}
\def\third{{\frac{1}{3}}}

\def\boxit#1{\vbox{\hrule\hbox{\vrule\kern3pt\vbox{\kern3pt#1\kern3pt}
    \kern3pt\vrule}\hrule}}

\def\sbar{{\langle s \rangle}}
\def\sigth{{ \sigma_{\theta} }}

\def\vareps{{\varepsilon}}

\def\Gambar{{\overline{\Gamma} }}

\def\bvec{{\bf b}}
\def\beffvec{{\bf b_{\rm eff}}}
\def\beff{{b_{\rm eff}}}
\def\bd{{b_e}}
\def\biso{{b_{iso}}}
\def\bisovec{{\rm b_{iso}}}
\def\bvecij{{\bvec}}

\def\Gamg{{\Gamma_g }}
\def\Gams{{\Gamma_s }}
\def\Gameps{{\Gamma_{\vareps}}}
\def\Gamepsave{{\langle \Gamma_{\vareps}\rangle}}
\def\rsvec{{\bf r_s}}
\def\rs{{r_s}}
\def\drsvec{{\bf \delta r_s}}
\def\drs{{\delta r_s}}

\def\drsveciso{{\bf \delta r_{s,iso}}}
\def\drsiso{{\delta r_{s,iso}}}
\def\rvec{{\bf r}}

\def\Qiss{{Q_{\rm ISS}}}
\def\Qpsr{{Q_{\rm PSR}}}
\def\Qnoise{{Q_{\rm NOISE}}}

\def\gigj{{g_i g_j^* }}
\def\ninj{{n_i n_j^* }}
\def\gigjave{{\langle \gigj \rangle}}
\def\Aave{{\langle {\rm A} \rangle}}
\def\Iave{{\langle {\rm I} \rangle}}
\def\Mave{{\langle {\rm M} \rangle}}
\def\Nave{{\langle {\rm N} \rangle}}
\def\Niave{{\langle {\rm N}_i \rangle}}
\def\Njave{{\langle {\rm N}_j \rangle}}
\def\ninjave{{\langle n_i n_j^* \rangle}}
\def\Irsvec{{{\rm I}_{\rsvec}}}

\def\Gi{{G_i}}
\def\Giave{{\langle G_i \rangle}}

\def\Gjave{{\langle G_j \rangle}}
\def\GiGjave{{\langle G_i G_j)\rangle }}
\def\Gave{{\langle G \rangle }}
\def\G2ave{{\langle G^2 \rangle }}
\def\sigC{{\sigma_C}}
\def\sigX{{\sigma_X}}

\def\fgam{{f_{\gamma}}}
\def\niss{{N_{\rm ISS}}}
\def\ndof{{N_{\rm dof}}} 
\def\Gbar{{{G}}}
\def\gbar{{\overline{g}}}

\def\Gambarmag{{\vert \Gambar \vert}}

\def\ths{{\theta_s}}
\def\thd{{\theta_d}}
\def\thij{{\theta_{ij}}}
\def\thiso{{\theta_{iso}}}

\def\Gij{{G}}
\def\Gijbar{{\overline{G}}}
\def\tp{{t^{\prime}}}
\def\tpp{{t^{\prime\prime}}}
\def\miss{{m_{\rm ISS}}}
\def\misssq{{m_{\rm ISS}^2}}
\def\misssqinv{{m^{-2}_{\rm ISS}}}

\def\mcross12{{m_{\rm cross}^2}}

\def\Dbar{{\overline{D}}}
\def\Ibar{{\overline{\rm I}}}

\def\gamg{{\gamma_{\rm g}}}
\def\gamG{{\gamma_{\rm G}}}

\def\rem{{r_{em} }}

\def\epsem{{\epsilon_{em}}}
\def\epsiso{{\epsilon_{iso}}}
\def\rlc{{r_{\rm LC}}}

\def\wdelta{{W_{\Delta}}}

\def\rbar{{\overline r}}
\def\Drmag{{\Delta r_{em}}}
\def\Drmagperp{{\Delta r_{em,\perp}}}

\tighten
\singlespace

\begin{document}

\title{Interstellar Seeing. I. Superresolution Techniques Using Radio
Scintillations}
\author{J. M.  Cordes}
\affil{\small Astronomy Department, 520 Space Sciences Building,
Cornell University, Ithaca, NY 14853  cordes@spacenet.tn.cornell.edu}

\vskip 0.1truein
\centerline{\bf Submitted to {\it Astrophysical Journal} \today}

\begin{abstract}

Interstellar scintillation 
can be used to probe transverse sizes of radio sources
on scales inaccessible to the nominal resolution of any terrestrial telescope,
e.g. $\lesssim 10^{-6}$ arc sec.
Methodology is presented that exploits this superresolution
phenomenon for both single aperture and interferometer observations. 
The treatment applies to the saturated (strong-scattering)
regime and holds for both  thin screens and extended media.
A general signal model for radio sources  is presented,
{\it scintillated amplitude modulated noise},
which applies to compact, incoherent synchrotron sources such as 
AGNs and gamma-ray
burst sources and also to known, coherent sources such as masers and pulsars.
The exact
probability density function for  measured intensities and
interferometric visibilities is obtained by solving a general Fredholm problem. 
An approximate density function is also obtained by using the equivalent
number of degrees of freedom in scintillation modulations.
The scintillation modulation variance is presented, which
includes the effects of source structure 
and time-bandwidth averaging in the signal processing. 
Two Bayesian methods are outlined for inferring the sizes of emission regions
that use first order statistics of the intensity and visibility.
Intensity cross-correlation methods for inferring source sizes are also given.
Intensity interferometry in the the radio context is compared 
to the optical intensity interferometry of Hanbury-Brown and Twiss.

\end{abstract}

\section{Introduction}
\label{sec:intro}

Diffractive interstellar scintillation (DISS) 
is caused by multipath scattering of radio
waves from small-scale density irregularities in the ionized interstellar
medium. It is sensitive to intrinsic sizes of radiation sources
in much the same way
that optical scintillation from atmospheric turbulence is quenched for
planets while strong for stars.    However, interstellar scintillation differs
from the atmospheric case in that it can resolve
sources  at angular resolutions much smaller than those
achievable with available apertures, including the longest baselines
used in very long baseline interferometry (VLBI), those using
space antennas.
Optical techniques such as
intensity interferometry,
speckle interferometry and adaptive optics
typically only {\it restore} the telescope resolution 
to what it would be in the absence of any atmospheric turbulence.

We define the {\it superresolution regime} where
the source is unresolved by terrestrial interferometers but is sufficiently
extended to modify the DISS.   Let
$\ths, \thij$ and $\thiso$ be the 
source size, interferometer fringe spacing, and isoplanatic DISS patch,
respectively.  
By definition, two point sources separated by much less
than the isoplanatic angle  will show identical DISS.
The isoplanatic angle $\thiso\sim \lambda / d\thd$, 
where 
$\lambda$ is the wavelength, 
$\thd$ is the size of the scattering (``seeing'') disk
and $d$ is the source-Earth distance. 
Using typical numbers
($d=1$ kpc, $\thd= 1$ mas at an observing frequency of 1 GHz),
$\thiso \sim 0.4\, \mu$arc sec.  For pulsars, whose light-cylinder
radii, $\rlc = cP/2\pi$ ($P$ = spin period) are smaller than
1 $\mu$arc sec at typical distances, we have   
$\thiso \lesssim \ths \ll \thij$  and $\thiso \ll \thd$.  
In this case, speckle methods can achieve far better resolution 
than the interferometer.  By speckle methods, we mean observations that
analyze differences in the DISS between source components, which
are sensitive to the spatial separations of those components.   
Cornwell \& Narayan (1993) have discussed particular superresolution 
techniques in the radio context.
In optical astronomy, superresolution is not achievable 
because $\thd \sim \thiso$.  However, the superresolution
regime has been identified in optical laboratory applications
(Charnotskii, Myakinin \& Zavorotnyy 1990).  

The purpose of this paper is to provide a rigorous and general formulation
of DISS that can be used in superresolution applications.  Previous work has
relied on DISS theories that are restricted to particular spatial geometries
(e.g. thin screens) or to particular wavenumber spectra (e.g. a power
law with the ``Kolmogorov'' slope.).  In subsequent papers, we will
apply the methods of this paper to scintillation observations
and derive constraints on pulsar source sizes.

Our treatment builds upon work published in both the optical and
radio propagation literature  and is unique in the following ways:
(1) it interfaces empirical astrophysical constraints on source radiation
fields with the scattering geometry and scattering strength appropriate
for radio observations; 
(2) it calculates fluctuation statistics of realistically
measureable quantities in single-aperture and interferometric
observations, taking into account all sources of fluctuation;
(3) we present exact calculations for the intensity 
probability density function (PDF) that take into
account arbitrary source brightness distributions and arbitrary amounts
of time-bandwidth averaging; and
(4) our results can be applied to a wide range of media with
arbitrary spatial extent and wavenumber spectrum. 
Gwinn \etal~ (1998) discuss issues that are very similar to those contained
in this paper.   Our treatment is more general than Gwinn \etal's because
it is not limited to scattering media contained in thin screens.
Also, our treatment of the probability distribution is not restricted
to small source sizes (compared to the isoplanatic scale).  Finally,
our treatment includes the effects of intrinsic source fluctuations,
which are assumed negligible by Gwinn \etal   

We restrict the analysis to strong (saturated) DISS.
Another component 
--- refractive interstellar scintillations (RISS) --- 
also modulates source intensitities, but on time scales much longer
than for DISS.   RISS can be seen from larger angular diameter sources,
by a factor of 1000, than can DISS.   We are interested in 
modeling data spans much shorter than the characteristic 
RISS time scale, so we ignore RISS in our discussion. 
The cases we consider are what would be called the ``single speckle''
regime in the optical literature.  This corresponds to the case where
the aperture size (either a single dish diameter or an interferometer
baseline) is smaller than the diffraction length scale 
(e.g. the Fried scale).  However, our formalism can be extended easily
to include aperture averaging and multiple speckle cases.

%In this paper we present a general signal model for scintillating sources
%and we derive statistical quantities that are sensitive to the source
%size, including the probability density function for the 
%scintillating flux density and its second moment.
%We derive the second moments of analogs to real-world measurements,
%e.g. the time-average intensity and visibility, taking into account
%intrinsic signal fluctuations and additive noise, as well as DISS.

In \S\ref{sec:history} we briefly summarize previous work on use of DISS
to resolve radio sources.
In \S\ref{sec:visextend} we 
present a general signal model and in \S\ref{sec:resolve}
give expressions for the modulation index
of the interferometric visibility and single aperture intensity. 
We take into account time-bandwidth
averaging and source extent.   
We calculate an approximate PDF of the visibility and intensity 
using the number of degrees of freedom in the scintillations.
We also outline the exact calculation 
of the PDF.  \S\ref{sec:bayes} presents a Bayesian inference method for
the source size. 
The paper is summarized in
\S\ref{sec:summary}.   The details of our definitions and calculations are
given in three Appendices.   
In Appendix ~\ref{sec:appendixa} we derive the statistics and 
scintillations of the scintillating amplitude modulated noise model.  
In Appendix ~\ref{sec:appendixb} we derive second moments for the intensity
and visiblity.  
In Appendix ~\ref{sec:appendixc} we derive the PDFs for the scintillations,
total intensity, and visiblity.   

Notation required in discussions of wave propagation through random media is
necessarily copious.  Apart from standard definitions
for wavelength, frequency, speed of light, and wavenumber
($\lambda, \nu, c$ and $k=2\pi/\lambda$, we list in Table 1 those 
symbols that are used throughout the paper. 
 
\section{Previous Applications of DISS Superresolution}
\label{sec:history}

DISS superresolution  techniques have been used in several ways to 
probe source structure. 
DISS has been sought from various kinds of AGNs 
(Condon \& Backer 1975; 
Armstrong, Spangler \& Hardee 1977;  
Condon \& Dennison 1978; 
Dennison \& Condon 1981), leading to bounds on the brightness 
temperature and limits on the Lorentz factors of bulk relativistic
flow.  
Lovelace (1970) first suggested that DISS might resolve pulsar
magnetospheres. 
Backer (1975) placed coarse limits on the sizes of pulsar 
magnetospheres by considering the fractional modulation of DISS
for a few pulsars.    Cordes, Weisberg \& Boriakoff (1983; hereafter CWB83)
 placed upper bounds on the source extent, and emission altitude, for
two pulsars, finding upper bounds on emission altitudes 
of $0.5\rlc$ and 0.1$\rlc$.   Wolszczan \& Cordes (1987) exploited
a remarkable episode of multiple imaging of a pulsar by
a refracting interstellar structure; they found that 
(a) different pulse components from a long-period pulsar
showed nonidentical DISS, signifying source resolution; 
and 
(b) the implied emission altitude is comparable to $\rlc$, in
contrast to other estimates, based on pulse widths and polarization,
which suggest emission altitudes of only 1-10\% of $\rlc$
(Blaskiewicz \etal~ 1991).
Kuzmin (1992) and Smirnova,  Shishov \& Malofeev (1996) 
reported similar results
on four additional long-period pulsars, again with
implied emission radii $\sim \rlc$.   

Recently, Gwinn \etal~ (1997,2000) have analyzed VLBI observations
of the Vela pulsar
using time and frequency resolutions that exploit the same spatial
resolving power of  DISS as do the single-dish observations
used by others.  
Through estimation of the probability density function (PDF) of
the visibility function magnitude, they infer that 
DISS shows  less modulation than expected from a point source
and therefore conclude that the source must be extended.  They 
estimate a transverse size 
$\sim 500$ km for the region responsible for  
the pulsed flux in a narrow range of pulse phase.
  
In a second paper, we 
reassess the conclusions of Gwinn \etal~ (1997) 
by considering how time-frequency averaging in the signal processing
affects inferences on source size from visibility fluctuations.
Gwinn \etal~ (1999) have also considered time-frequency averaging and the
effects of source fluctuations.  The approaches differ and yield 
different conclusions about the importance of averaging and source noise
as well as differing on estimates of the source size.

Also recently, radio observations of gamma-ray burst (GRB) afterglows
suggest that DISS occurs in the  early stages and then is 
quenched as expanding synchrotron sources are first smaller than,
and then exceed, the isoplanatic scale (Goodman 1997; Frail \etal~ 1997).
If GRB synchrotron sources are incoherent sources, the angular size
requirements for DISS to appear are severe.   
We defer to another paper a discussion of GRB scintillation.

\section{Signal Model for Scintillating Sources}
\label{sec:visextend}

To account for all contributions to measureable quantities, we need
a comprehensive statistical model for the received signal.  The model
presented here includes a source that is temporally incoherent but has
arbitrary spatial coherence; diffractive interstellar scintillation
from an arbitrary distribution of scattering material along the line of
sight; and additive radiometer  noise.

We define $\vareps(\rvec, t)$ to be the complex, narrowband, baseband
scalar electric field  that is explicitly or implicitly manipulated
in radio astronomy systems.  (For definitions, see Appendix A.)  
It is determined by the source emission mechanism and by
propagation effects through intervening media,   
as well as by receiver and background sky noise. 
We consider 
source emission that has underlying Gaussian statistics and 
propagation effects from the turbulent, ionized interstellar medium (ISM).  
The field measured at position $\rvec$ 
is the superposition of scintillating source
components and radiometer noise,
$n(\rvec, t)$,
\be
\vareps(\rvec,t) = \int d\rsvec\,
        \vareps_s(\rsvec, t) g(\rvec, t, \nu, \rsvec) + n(\rvec, t).
\label{eq:sigmod}
\ee
All vectors are two-dimensional and  transverse to the line of sight.
The corresponding intensity is
\be
I(\rvec, t) = \vert \vareps(\rvec, t)\vert^2.
\ee

The quantity $\vareps_s(\rvec, t)$ is the field emitted per unit area
at the source 
but whose amplitude includes implicitly an inverse distance dependence.
Propagation is described by the quantity
$g(\rvec, t, \nu, \rsvec)$, 
which is the propagator for a point source at location $\rsvec$.
It includes a phase factor for free-space propagation
as well as phase and amplitude factors associated with DISS.\footnote{
The form of Eq.~\ref{eq:sigmod} is approximate.  Factoring the integrand relies
on $\vareps_s$ being narrowband with bandwidth $\Delta\nu \ll \nu$, that
it varies much faster than the DISS propagator, $g$, and
that the bandwidth is much smaller than the 
characteristic scintillation bandwidth,  
i.e. the characteristic bandwidth on which $g$ changes.
This last constraint is
identical to requiring that differential propagation times 
be much less than the shortest characteristic time scale of
the signal.
Later, we also consider variations of $g$ with frequency.  The signal model
presented still applies if we consider the total frequency range to
comprise many separate intervals in each of which $g$ is piecewise constant.}
Here $\nu$ is the center frequency of the passband with bandwidth $\Delta\nu$
 that is selected by
the receiver and mixed to baseband.    The notation for $\vareps(\rvec, t)$
and $n(\rvec, t)$ leaves this center frequency implicit; but 
each of these quantities is a time series of a narrowband process
and is the baseband equivalent of the narrowband radiation field selected by
the receiver.
Further justification for this model is given in Appendix A.

The field emitted by the source  is taken to be amplitude modulated noise 
(Rickett 1975; Cordes 1976b), 
$\vareps_s(\rsvec, t) = a(\rsvec, t)  m(\rsvec, t)$,
which is a physically motivated and empirically confirmed model for
most astrophysical sources.
The amplitude modulated noise model includes 
nonstationary modulations $a(\rsvec, t)$  
of stationary noise $m(\rsvec, t)$.  
The noise correlation function is
$\langle m(\rsvec_1, t_1) m^*(\rsvec_2, t_2) \rangle = 
	\delta(\rsvec_2-\rsvec_1) \Delta(t_2 - t_1)$,
where the asterisk denotes complex conjugation,
angular brackets denote ensemble average, and $\Delta(\tau)$ is a 
continuous delta-function-like quantity with 
$\Delta(0) = 1$ and width equal to the reciprocal bandwidth of
the receiving system.  The additive noise, $n(\rvec, t)$, 
also is `$\Delta$' correlated in time and is assumed spatially  
uncorrelated across nonzero baselines, $\bvec = \rvec_2 - \rvec_1 $.    

\subsection{Interferometer Visibility Function \& Phase Structure Function}

It is well known that the mean visibility function of 
a scattered point source (in strong scattering) is
the product of the true source visibility $\Gams(\bvec)$ 
and 
the second moment of the DISS modulation, $\gamg$
(e.g. Rickett 1990 and references therein): 
\be
\langle \Gameps(\bvec, \taui) \rangle  = 	
\langle \vareps(\rvec, t) \vareps^*(\rvec+\bvec, t+\taui) \rangle  = 
			\Delta(\taui) 
			\gamg(\bvec, 0, 0, 0)
			\Gams(\bvec).
\label{eq:visfn}
\ee
The source visibility is the usual Fourier transform of the 
brightness distribution, 
$I_s(\rsvec) = A(\rsvec) = a^2(\rsvec)$,
\be
\Gams(\bvec) = 
	\int d\rsvec \, e^{+ik\dpo^{-1}\rsvec\cdot\bvec}
	I_s(\rsvec),
\label{eq:vis}
\ee
where $d$ is the source-observer distance and  $k=2\pi c^{-1}\nu$.
We use a spatial vector, $\rsvec$, to define
the source brightness distribution rather than using an angular variable, as is 
common practice.  
We assume that the  propagation delay between the
pair of  sites has already been removed, so the visibility function
maximizes at $\taui \lesssim (\Dnu)^{-1}$.   

The factor $\gamg$ in Eq.~\ref{eq:visfn}  is proportional to the 
second cross moment of the propagator, $g(\rvec, t, \nu, \rsvec)$,  at two
observation positions,  
times and 
frequencies
separated by 
$\bvec$, 
$\tau$, 
$\delta \nu$,
respectively, 
and for two point sources separated by $\delta\rsvec$:
\be
\Gamg(\bvec, \tau, \delta\nu, \delta\rsvec) &=& 
\langle g(\rvec, t, \nu, \rsvec) 
      g^*(\rvec+\bvec, t+\tau, \nu+\delta\nu, \rsvec+\delta\rsvec) \rangle 
	= e^{-i\psi} \gamg(\bvec, \tau, \delta\nu, \delta\rsvec).  
\label{eq:gamggeneral} 
\ee
In this equation, the phase $\psi$ is determined 
by free-space propagation  and drops out in much of what
appears below, but is responsible for the Fourier relation in 
Eq.~\ref{eq:vis}.  Appendix ~\ref{sec:appendixa}  shows the details. 
The form for $\gamg$ with zero frequency lag ($\delta\nu=0$) is 
simply expressed
in the Gaussian limit using the phase structure function,
$\dphi(\bvec, \tau,  \drsvec)$,
\be
\gamg(\bvec, \tau, \delta\nu =0, \delta\rsvec) &=& 
e^{-\half \dphi(\bvec, \tau,  \drsvec)}. 
\label{eq:gamg} 
\ee
For nonzero frequency lags, a closed-form expression for $\gamg$ is
not usually available.
The DISS ``gain''
$G = \vert g \vert^2$,  
has unit mean, $\langle G \rangle = 1$, and
has a normalized autocovariance in the  strong scattering 
(Rayleigh) limit, 
\be
\gamG (\bvec, \tau, \delta\nu, \drsvec) = 
\left\langle
G(\rvec, t, \nu, \rsvec)G(\rvec+\bvec, t+\tau, \nu+\delta\nu, \rsvec+\delta\rsvec)
\right\rangle
-1 = 
\vert \gamg(\bvec, \tau, \delta\nu, \drsvec)\vert^2. 
\label{eq:gamG}
\ee
For a medium in which scattering occurs with variable
strength all along the line of sight, 
the phase structure function is
(Lotova \& Chashei 1981; Cordes \& Rickett 1998) 
\be
\dphi(\bvec, \tau, \drsvec) &\propto& 
	\lambda^2 \int_0^d ds\, \cnsq(s) \, \vert \beffvec(s) \vert^{\alpha} 
   \label{eq:dphi}\\ 
\beffvec(s)  &=& (s/d) \bvec + \veffvec(s)\tau + (1-s/d) \drsvec 
   \label{eq:beffvec}\\
\veffvec(s) &=& (s/d)\vobsvec + (1-s/d)\vpvec - \vismvec(s). 
   \label{eq:veffvec}
\ee
$\cnsq$  is the coefficient in the wavenumber spectrum for electron
density variations  and $\alpha$ is the exponent of the structure function.  
For a square-law structure function $\alpha = 2$, while for a Kolmogorov
medium in strong,  but not superstrong scattering 
(Cordes \& Lazio 1991), $\alpha = 5/3$.
$\vpvec$ is the pulsar velocity, $\vobsvec$ is the observer's velocity
and $\vismvec$ is the velocity of the scattering material in the ISM.
For the case $\cnsq(s) \propto \delta(s-\ds)$, where $\ds$ is the distance
of a scattering screen from a pulsar,  
we retrieve the result applicable for a thin-screen.

Note that the phase structure function $\propto \lambda^2$ for radio
propagation through tenuous plasmas.  For 
optical and infrared (IR) propagation through the atmosphere,
$\dphi \propto \lambda^{-2}$. 

Our expressions Eq.~\ref{eq:gamG}-\ref{eq:veffvec} are quite general,
being based on Gaussian statistics for the wavefield and on the
saturated (Rayleigh) regime of scattering.  As such, they can be used
for observations of Galactic and extragalactic radio sources,
including pulsars, masers, microquasars, active-galactic nuclei,
and gamma-ray burst sources.

%\subsubsection{Anisotropic Scattering}

Our form for $\dphi$ applies to isotropic scattering irregularities.
Evidence exists for anisotropies in heavily scattered sources
(Frail \etal~ 1994; 
Wilkinson, Narayan \& Spencer 1994;
Yusef-Zadeh 1994; 
Molnar \etal~ 1995;
Desai \& Gwinn 1998; 
Spangler \& Cordes 1998;
Trotter, Moran \& Rodriguez 1998).
Though, for simplicity, we consider only the isotropic case in this paper,
it is a simple matter to extend our results to the anisotropic  case,
which we will do elsewhere.

\subsection{Isoplanatic Scales} \label{sec:iso}

DISS is correlated over spatial and temporal scales at the observer's 
location that are determined by contours of constant $\dphi$.
For a thin screen at $s=\ds$ from the source, we write
$\dphi = (\vert\beffvec\vert / b_e)^{\alpha}$, with $\beffvec$ given by 
Eq. \ref{eq:beffvec} and where $b_e$ is the 1/e scale of the structure function.
 We define the isoplanatic length scale
and time scale through 
$\dphi(\bisovec, 0, 0) = 1$ and
$\dphi(0 , \dtd, 0) = 1$,
yielding $b_{iso} = b_e (d/\ds)$
and, if the source's  speed dominates $\veff$ (as it does for many pulsars),
$\dtd = b_e /  (1-\ds/d) \vp$.  
We also define the isoplanatic scale at the source's location
using $\dphi(0, 0, \drsveciso) = 1$, resulting in  
$\drsiso = b_e ( 1 - \ds/d)^{-1} = \vp \dtd$.  
The isoplanatic scale $\drsiso$ defines the separation at which
two point sources would scintillate with a correlation coefficient
of $e^{-1}$.  This scale determines whether DISS can resolve a source,
as discussed in the Introduction.    
For reference, 
the Fried scale $r_0$,  defined in the optical and IR literature,
is related to our definitions using
$\dphi(b, 0, 0) = 6.88 (b/r_0)^{5/3}$ (e.g. Goodman 1985),
so $r_0 \approx 3.2 b_e$.
Also, in the radio case $r_0\propto \lambda^{-6/5}$ while
$r_0 \propto \lambda^{+6/5}$ for optical/IR propagation.

In strong scattering, the isoplanatic scale is smaller, in some cases by
three orders of magnitude or more, than the Fresnel scale.  The Fresnel
scale, at meter wavelengths, $\sim \sqrt{\lambda D} \approx 10^{11}$ cm
for kiloparsec distances. 

The isoplanatic scale $\drsiso$ is smaller for sources that
are scattered more heavily.  This corresponds to more distant sources,
sources viewed through regions of excess scattering, or sources observed
at longer wavelengths.    For a continuous source, such as one with a
Gaussian brightness distribution, its size  compared to $\drsiso$ determines
the depth of modulation of the DISS.  To apply this basic idea, however,
averaging over time and over the receiver bandwidth must also be dealt
with carefully because it too affects the depth of modulation.   We now
consider all these effects in what follows.

\section{Resolving Sources with Scintillations} \label{sec:resolve}

Several methods can be used to exploit the scintillation 
phenomenon in order to resolve sources.  These are
(1) measurement of the fractional modulation of the source through analysis
of the intensity variance;
(2) estimation of the intensity PDF, equivalent to analysis of {\it all}
moments;
and (3) use of cross-correlation functions for the DISS of separate sources
to measure the spatial offsets of those sources.

\subsection{Visibility and Intensity Statistics}\label{sec:stats}

The visibility function and the intensity are both second field moments.
Estimates  of these second moments from finite data sets fluctuate by
amounts that are formally described by the fourth field moment.  Encoded
in these fluctuations is information about source structure and the
intervening medium.   We use a normalized fourth moment --- a generalized
modulation index (squared) ---
to cast intensity and visibility fluctuations in 
a similar form.    The modulation index includes
the effects of averaging over time and frequency
and is calculated for an arbitrary source brightness distribution.  
 
\subsection{Autocorrelation Functions}

To model realistic cases, we take into account averaging over 
time and the finite bandwidth of the narrowband signal.  For simplicity,
we refer to the finite bandwidth as ``frequency averaging.'
To resolve DISS, the averaging intervals $T$ and $B$ 
must be smaller than the characteristic correlation scales of
$G$, the DISS gain in time and frequency.  These are usually called
the DISS or `scintillation'  time scale and bandwidth,   
denoted $\dtd$ and $\dnud$, respectively.

Let $\Ibar(\rvec, t)$ be the intensity at location $\rvec$
calculated for a narrowband signal with bandwidth $B$
after averaging over the interval
$[t - T/2, t + T/2]$.
Define also the averaged visibility function as
 $\Gambar(\bvec, t)$, calculated between two sites separated by
baseline $\bvec$ and with time lag $\tau$ but with zero 
frequency lag.
The autocorrelation functions (ACFs) of the intensity and visibility
are
\be
R_{\Ibar}(\bvec, \taubar) \equiv  
	\left\langle \Ibar(\rvec, t) \Ibar(\rvec+\bvec, t+\taubar) \right\rangle 
\label{eq:RIbarmain}
\ee
\be
R_{\Gambar}(\bvec, \taubar) \equiv
	\left\langle \Gambar(\rvec, t) \Gambar^*(\rvec+\bvec, t+\taubar) \right\rangle 
\label{eq:RGambarmain}
\ee

\subsection{Modulation Indices}\label{sec:modindex}
To quantify fluctuations of $\Ibar$ and
$\Gambar$,  we calculate the modulation indices, 

\be
m^2_{\Ibar}(\bvec, \tau) &\equiv&
	\frac{\displaystyle{ R_{\Ibar}(\bvec, \tau) - \Iave^2}}{\Iave^2} \\  
m^2_{\Gambar}(\bvec, \tau) &\equiv&
	\frac{\displaystyle{ R_{\Gambar}(\bvec, \tau) - 
		\vert\Gamepsave\vert^2}}{\Iave^2},  
\ee 
where we normalize by the mean intensity in both cases.
As shown in Appendix ~\ref{sec:appendixb}, the modulation index receives
contributions from four terms.  For compact sources, 
the dominant term is in fact caused by DISS and is given by
\be
\misssq(\bvec, \taubar) &=& 
	  \langle I \rangle^{-2}   
		\int\int d\rsvec_1\, d\rsvec_2\, 
		I_s(\rsvec_1)\,
		I_s(\rsvec_2)\, 
		\Qiss(\bvec, \taubar, \rsvec_2 - \rsvec_1, T, B),
\label{eq:miss2}
\ee
where, for uniform averaging\footnote{Uniform averaging
over frequency corresponds to spectrometer passbands that are 
perfectly rectangular;  real passbands, $h(\nu)$, have shapes similar to
Gaussian functions 
and can be handled by replacing $1-\vert \delta\nu/B \vert$ in Eq.~\ref{eq:Qdef}
with the autocorrelation of $h(\nu)$ and extending the 
limits to $\pm\infty$.  Computations show
that this refinement produces no significant changes in our discussion.}
in $T$ and $B$,
\be
\Qiss(\bvec, \taubar, \delta\rsvec, T, B) 
&=& 
(TB)^{-1}
     	\int_{-T}^{+T} d\tau^{\prime} 
		\left(1 - \left\vert \frac{\tau^{\prime}}{T} \right\vert \right) 
	\int_{-B}^{+B} d\delta\nu\, 
		\left(1 - \left\vert \frac{\delta\nu}{B} \right\vert \right)
\nonumber\\
&~&\quad \quad\quad 
	\times
	\left \{  
	\begin{array}{ll}
	   	  \gamG(\bvec, \tau^{\prime}+\taubar, \delta\nu, \delta\rsvec) 
		& \mbox{{\rm \, intensity}} \\
\\
		\vert \Delta(\taui)\vert^2 \,
		e^{-ikd^{-1}\bvec\cdot\delta\rsvec} 
	   	  \gamG(0, \tau^{\prime}+\taubar, \delta\nu, \delta\rsvec) \quad 
		& \mbox{{\rm \, visibility}}, 
	\end{array}
	\right.
\label{eq:Qdef}
\ee
with $\gamG$ defined in Eq.~\ref{eq:gamG}. 
Visibility fluctuations are independent of baseline $\bvec$ for unresolved
sources, for which the complex exponential $\to 1$, while 
intensity variations depend more strongly on $\bvec$, even for 
sources unresolved by the baseline.  The baseline-independent
property for visibility fluctuations
 is similar to the conclusion found by Goodman \& Narayan (1989).  
The difference between intensity and visibility statistics arises from
the ordering of the time-averaging and cross-correlation 
operations in the two cases. 

The modulation index, as presented here, includes only DISS fluctuations.
There are additional contributions to the 
visibility or intensity variance from intrinsic source fluctuations
and from additive noise.  These are secondary to our discussion here,
but are important in any practical application where the time-bandwidth
product is low and where intrinsic source fluctuations are high.
 For pulsars, pulse-to-pulse amplitude
variations are important when only a few pulses are included in any
averaging. 
Appendix ~\ref{sec:appendixb} gives full expressions for all
contributions  to intensity variations.
 
\subsection{Number of Degrees of Freedom in Fluctuations}\label{sec:ndof}

TB averaging and extended structure 
expressed in the integrals of $\gamG$ 
in Eq.~\ref{eq:miss2},\ref{eq:Qdef} 
diminish scintillation fluctuations.
The modulation index of the averaged intensity or
visibility  depends on
time averaging and source extension in similar ways  because both
increase the number of degrees of freedom 
in the integrated intensity.
The number of degrees of freedom is 
\be 
\ndof = 2\misssqinv = 2\niss \ge 2,
\label{eq:ndof}
\ee
where $\niss$ is the number of independent DISS fluctuations 
(``scintles'') that are averaged.  For observations in the 
single speckle regime, where scintles are resolved in time and frequency,
we expect $1 \le \niss \lesssim 2$.  

\subsection{Examples}

\begin{figure}
\plotfiddle{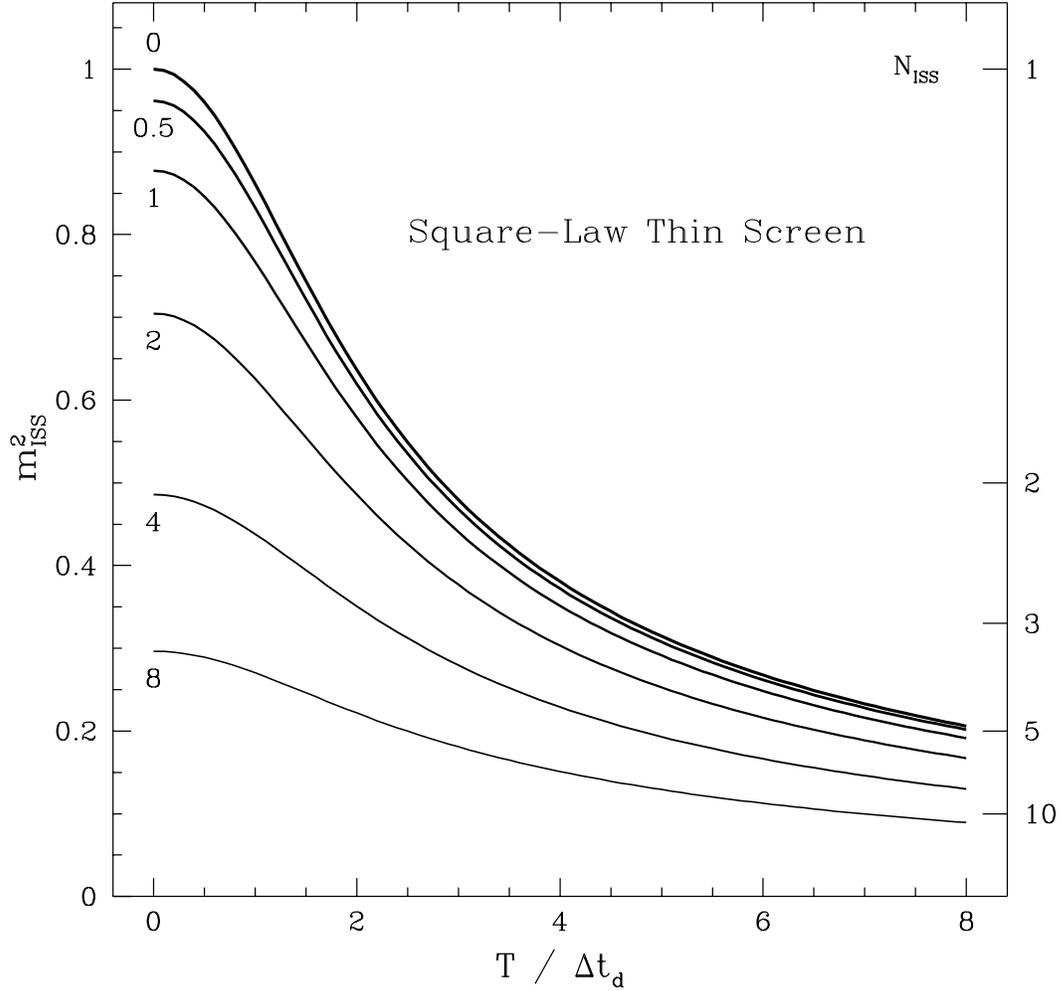}{4.5truein}{0}{70}{70}{-230}{-80}
\caption{
The DISS modulation index plotted against averaging time in
units of the characteristic diffraction time, $T/\dtd$ for 
values of the frequency resolution in units of the characteristic
diffraction bandwidth, $B/\dnud = 0, 0.5, 1, 2, 5, 10$.
The right hand scale gives the effective number of ISS fluctuations
that are averaged, $\niss = 1/\misssq$.
This case applies to a thin-screen having a square-law structure function.
}
\label{fig:missSqLaw}
\end{figure}

\begin{figure}
\plotfiddle{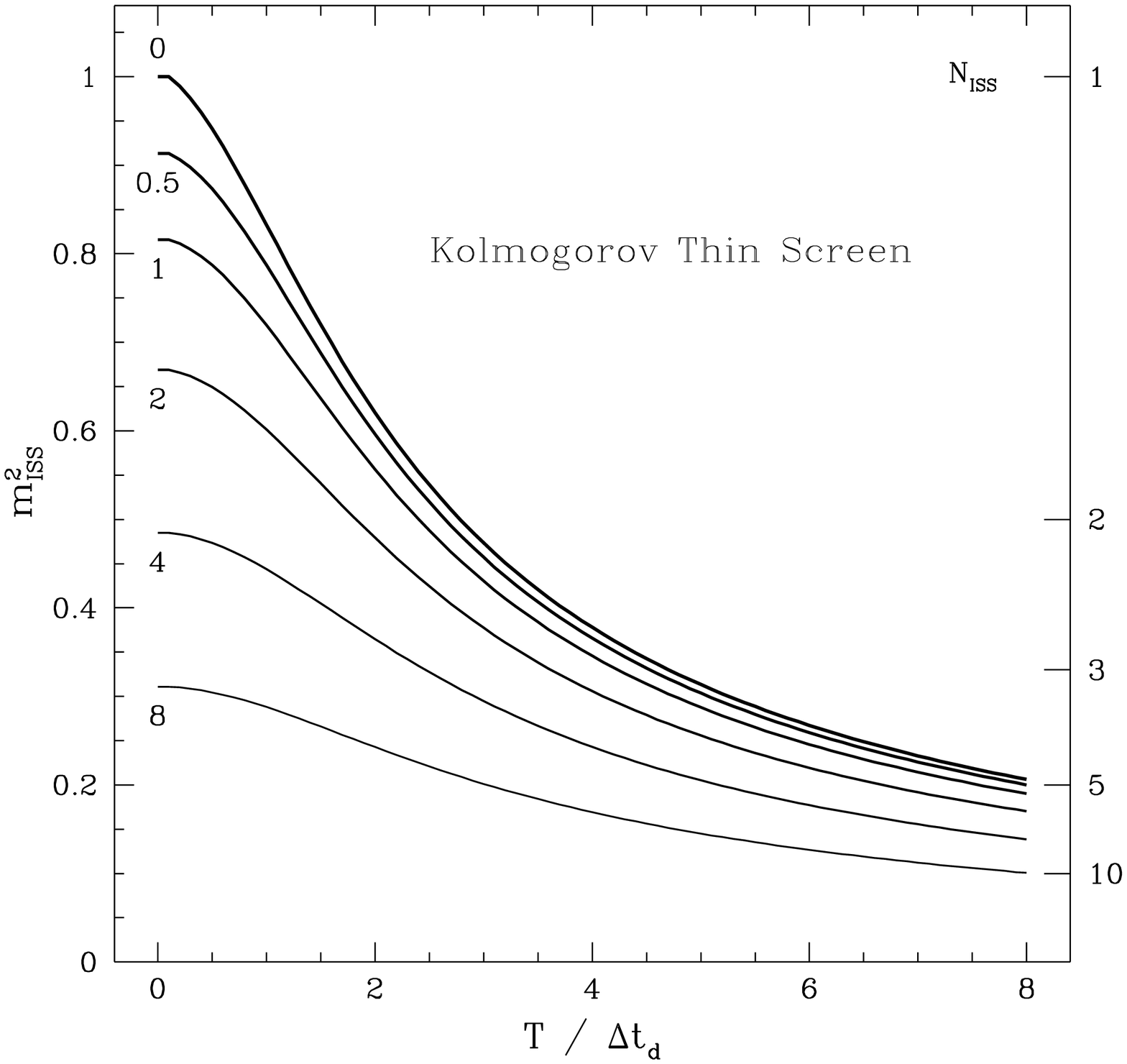}{4.5truein}{0}{70}{70}{-230}{-80}
\caption{
The DISS modulation index plotted against averaging time in
units of the characteristic diffraction time, $T/\dtd$ for 
values of the frequency resolution $B$ in units of the characteristic
diffraction bandwidth, $B/\dnud = 0, 0.5, 1, 2, 4, 8$,
as labelled.
This case is for a thin screen with  a Kolmogorov wavenumber 
spectrum.  The right-hand scale indicates the number of ISS
fluctuations (`scintles') averaged, $\niss$. 
}
\label{fig:missKols}
\end{figure}

\begin{figure}
\plotfiddle{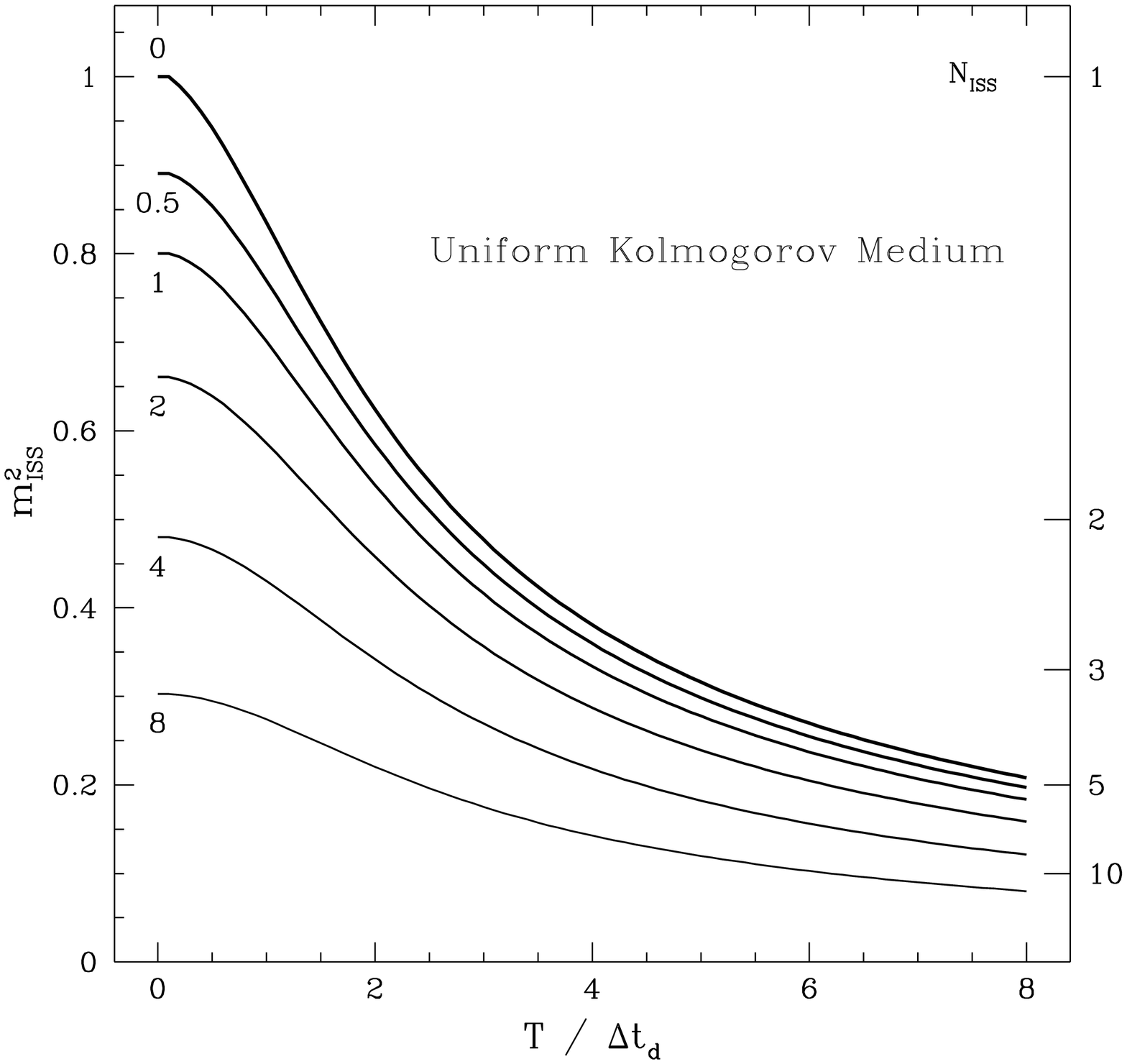}{4.5truein}{0}{70}{70}{-230}{-80}
\caption{
The DISS modulation index plotted against averaging time in
units of the characteristic diffraction time, $T/\dtd$ for 
values of the frequency resolution $B$ in units of the characteristic
diffraction bandwidth, $B/\dnud = 0, 0.5, 1, 2, 4, 8$,
as labelled.
This case is for a uniform medium  with  a Kolmogorov wavenumber 
spectrum.  The right-hand scale indicates the number of ISS
fluctuations (`scintles') averaged, $\niss$. 
}
\label{fig:missKolu}
\end{figure}

In Figures \ref{fig:missSqLaw}-\ref{fig:missKolu} we show 
$\misssq$ plotted against integration time $T$ for different values of
bandwidth $B$.   These cases are for a point source.
In the figures, we use $T$ and $B$ normalized by
the scintillation time scale, $\dtd$, and bandwdith, $\dnud$. 
Figure~\ref{fig:missSqLaw} is the case for a thin screen with a 
square-law structure function (i.e. $\alpha = 2$).
Figure~\ref{fig:missKols} is for a thin screen with $\alpha=5/3$,
the form appropriate for  a Kolmogorov medium if the scattering is
strong but not ultrastrong (Cordes \& Lazio 1991).  Finally,
Figure~\ref{fig:missKolu} is the Kolmogorov case for
an extended medium with uniform statistics along the line of sight.

The differences between the plotted curves for the different media
are subtle, but nonetheless significant if one were to use the
modulation index (for given $T,B$, say) to try to determine what kind
of medium was responsible for a given measurement.   More importantly,
the differences must be considered when assessing whether the source
is extended.    

To consider source-size effects,
we adopt a circular Gaussian brightness distribution with
source size, $\sigma_r$, 
\be
I_s(\rsvec) = 
	\left ( 2\pi\sigma_r^2 \right )^{-1}
	\exp 	\left ( - \frac {\vert\rsvec \vert^2}  {2\sigma_r^2}
		\right ).
\label{eq:gaussbright}
\ee
For this case, the squared modulation index is
\be
\misssq(\bvec, \taubar) &=& 
		(4\pi\sigma_r^2)^{-1}
		\int d\delta\rsvec\, 
		e^{-(\vert\delta \rsvec \vert /2\sigma_r)^2}
		\Qiss(\bvec, \taubar, \delta\rsvec, T, B).
\label{eq:miss2-gaussian}
\ee

\begin{figure}
\plotfiddle{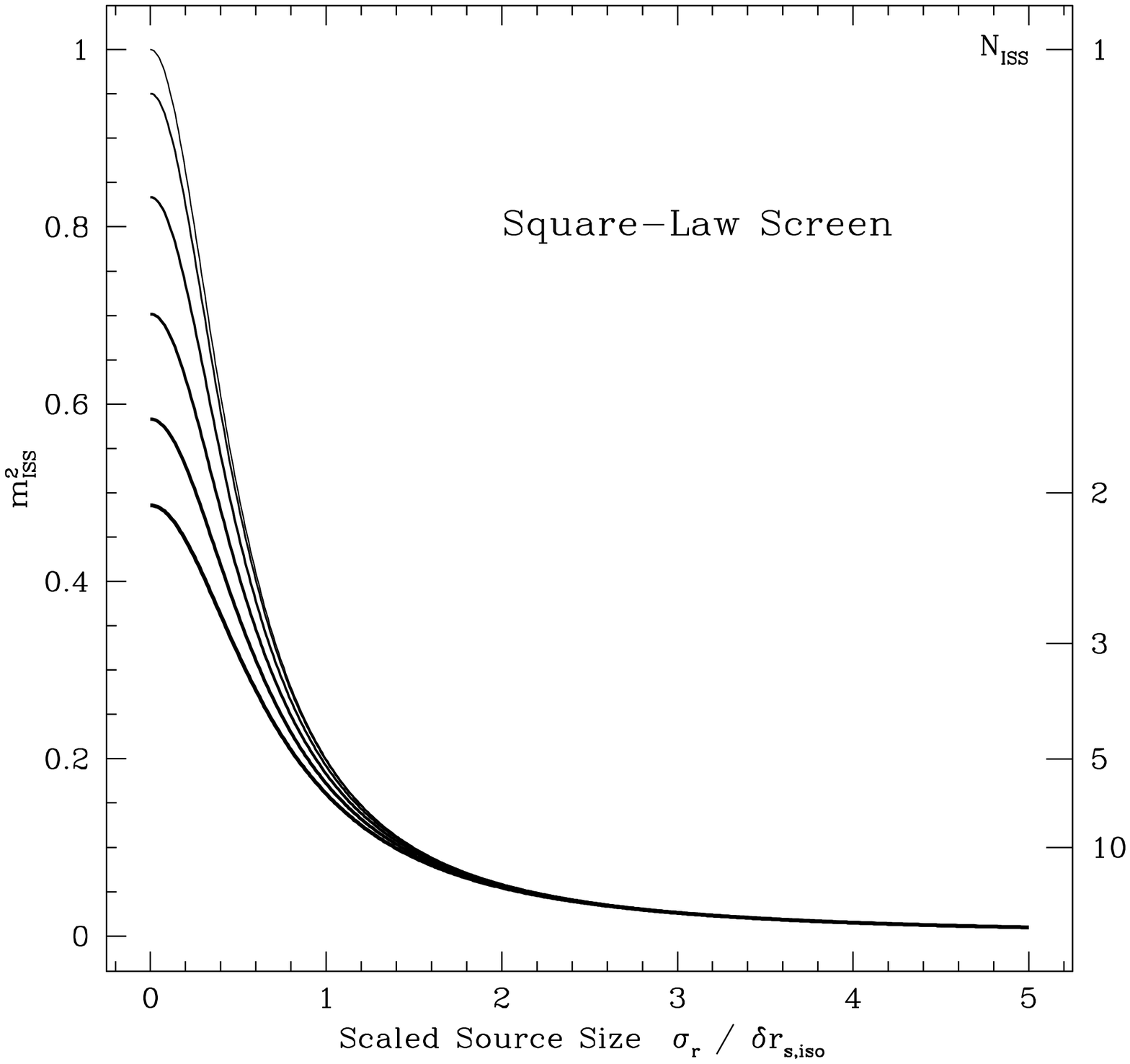}{4.5truein}{0}{70}{70}{-230}{-80}
\caption{
The DISS modulation index plotted against source size in
units of the isoplanatic scale, $\drsiso$.   The different curves are
for different values of averaging time $T$ and bandwidth $B$
in units of the characteristic
diffraction time scale and bandwidth. 
In order of light to heavy lines, the curves are
$(T/\dtd, B/\dnud) = 
(0,0),
(0.4, 0.4),
(0.8, 0.8),
(1.2, 1.2),
(1.6, 1.6),
(2.0, 2.0)$.
This case is for a thin screen with  a square-law structure function. 
The right-hand scale indicates the number of ISS
fluctuations (`scintles') averaged, $\niss$. 
}
\label{fig:miss-vs-ss-sqlaw}
\end{figure}

\begin{figure}
\plotfiddle{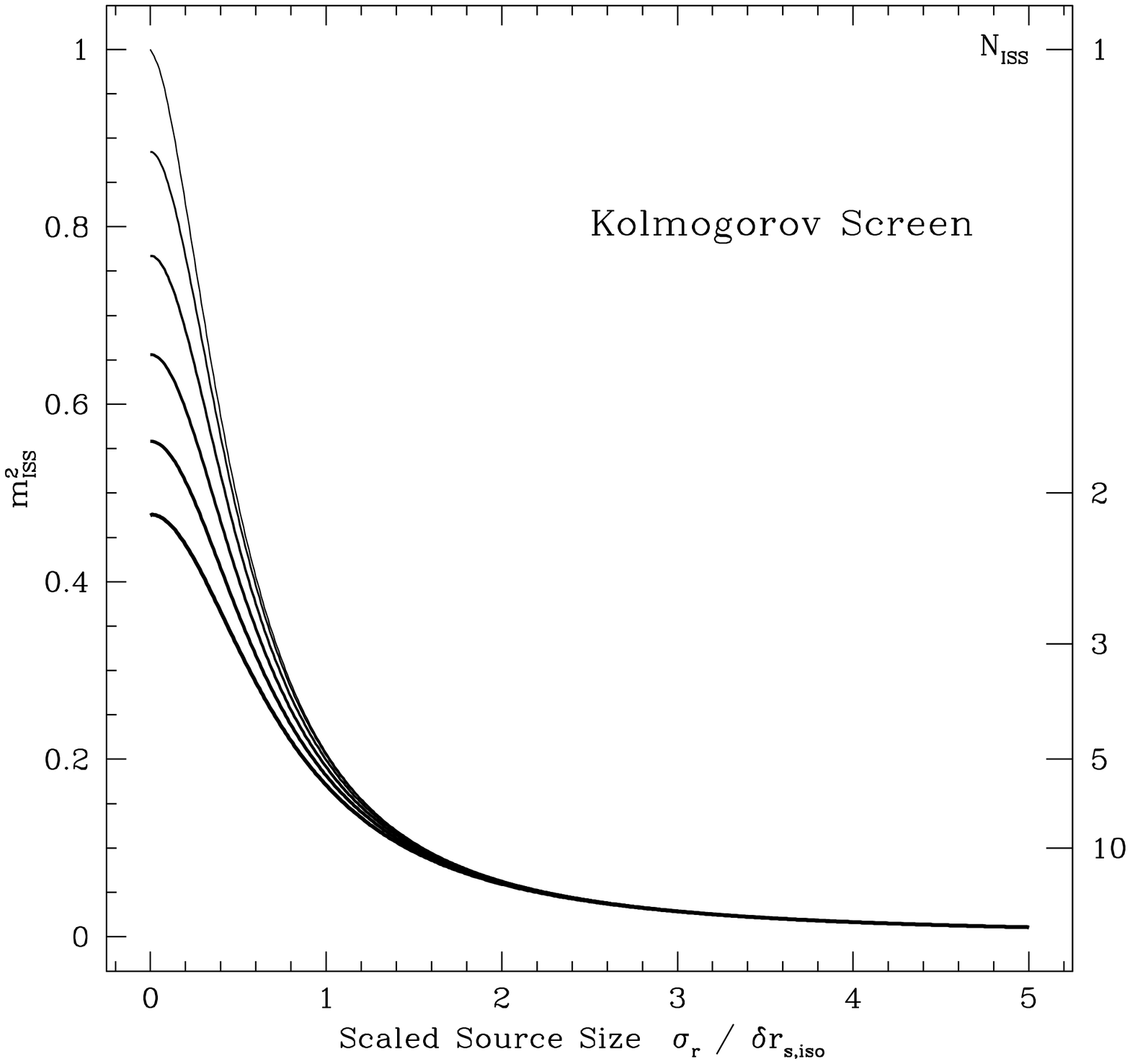}{4.5truein}{0}{70}{70}{-230}{-80}
\caption{
The DISS modulation index plotted against source size in
units of the isoplanatic scale, $\drsiso$.   The different curves are
for different values of averaging time $T$ and bandwidth $B$
in units of the characteristic
diffraction time scale and bandwidth. 
In order of light to heavy lines, the curves are
$(T/\dtd, B/\dnud) = 
(0,0),
(0.4, 0.4),
(0.8, 0.8),
(1.2, 1.2),
(1.6, 1.6),
(2.0, 2.0)$.
This case is for a thin screen with  a Kolmogorov
($\alpha=5/3$)  structure function. 
The right-hand scale indicates the number of ISS
fluctuations (`scintles') averaged, $\niss$. 
}
\label{fig:miss-vs-ss-Kol-s}
\end{figure}

\begin{figure}
\plotfiddle{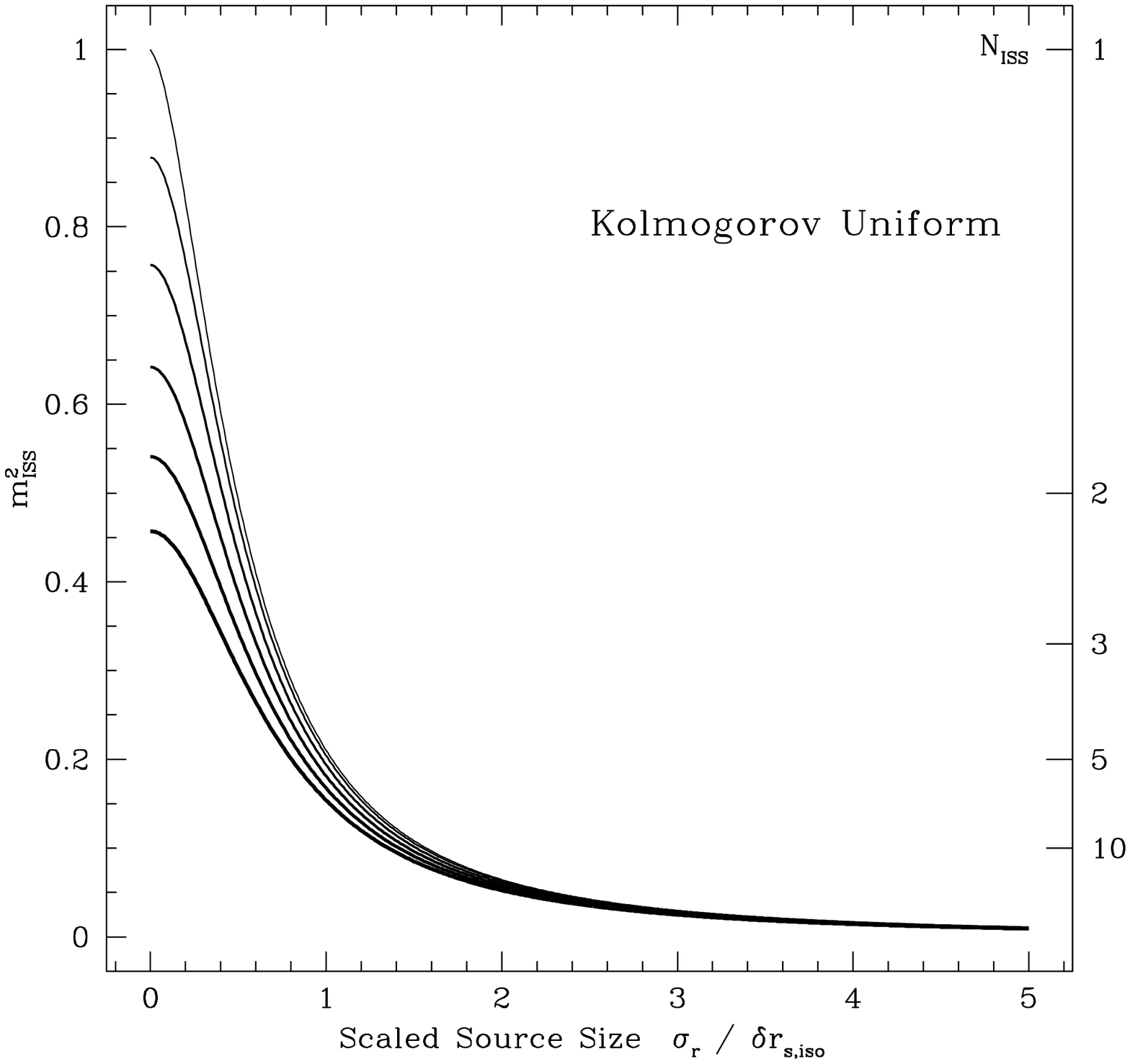}{4.5truein}{0}{70}{70}{-230}{-80}
\caption{
The DISS modulation index plotted against source size in
units of the isoplanatic scale, $\drsiso$.   The different curves are
for different values of averaging time $T$ and bandwidth $B$
in units of the characteristic
diffraction time scale and bandwidth. 
In order of light to heavy lines, the curves are
$(T/\dtd, B/\dnud) = 
(0,0),
(0.4, 0.4),
(0.8, 0.8),
(1.2, 1.2),
(1.6, 1.6),
(2.0, 2.0)$.
This case is for a uniformly extended medium 
with a Kolmogorov  ($\alpha=5/3$) structure function. 
The right-hand scale indicates the number of ISS
fluctuations (`scintles') averaged, $\niss$. 
}
\label{fig:miss-vs-ss-Kol-u}
\end{figure}

Figures \ref{fig:miss-vs-ss-sqlaw}-\ref{fig:miss-vs-ss-Kol-u}
show $\misssq$ plotted against source size in units of the
isoplanatic scale, $\drsiso$ (defined in \S\ref{sec:iso}).
The three figures are for the same square-law, thin-screen Kolmogorov,
and uniform Kolmogorov media considered in Figures
\ref{fig:missSqLaw}-\ref{fig:missKolu}, which are results for point
sources.   The different curves are
for different combinations of $T/\dtd$ and $B/\dnud$.   One 
feature to note is that $\misssq$ falls off more rapidly with
source size $\sigma_r/\drsiso$ than it does with
$T/\dtd$ or with $B/\dnud$.   This is because the source-size dependence
results from a two-dimensional integration over the difference
vector $\rsvec_2 - \rsvec_1$ in Eq.~\ref{eq:miss2} as 
compared with one-dimensional integrals for TB averaging.   
These figures show again that any inference on source
size must account not only for TB-averaging, but also for the type of
medium underlying the measurements.    Furthermore, the predicted
contributions from TB-averaging rely on accurate measurements of the
DISS time scale and bandwidth.

\subsection{Interpretation of Modulation Index}

Application of Eq.~\ref{eq:miss2} is as follows.
If $\misssq = 1$ (within errors) then the source is unresolved 
by the DISS, the baseline $\bvec$ has not resolved the scattering
disk, {\it and} the scintillations cannot have decorrelated over
the averaging intervals $T$ and $B$.  The DISS gain $G$
then has an exponential
PDF associated with the two degrees of freedom in the scattered
wavefield. 

Alternatively, $\misssq < 1$ can signify 
(1) resolution of the scattering disk by the baseline (in the case
    of intensity interferometry);
(2) variation of the DISS over the averaging time or averaging bandwidth;
{\it or} that
(3) the source has been resolved by the DISS, i.e. that it is 
comparable to or larger than the isoplanatic scale of the 
DISS.  
To discriminate between these possibilities, auxiliary information  
is needed 
that characterizes the dependence of $\gamg$ 
on its four arguments, $\bvec, \tau, \delta\nu$, and $\delta\rsvec$. 
Such information is obtained by making DISS and angular broadening 
measurements over a wide range of frequencies (e.g. Rickett 1990).

Complications in estimating $\misssq$ arise from the fact that scintillating
sources fluctuate, on inverse-bandwidth time scales and on a variety of
longer time scales, and there is additive noise in any real-world receiver 
system.  We consider all such complications in Appendices 
\ref{sec:appendixb} and  \ref{sec:appendixc} and also in the next few sections. 

\subsection{PDF of the Averaged DISS Gain}

While the modulation index of visibility fluctuations may allow
inference of source structure, an analysis of the full probability
density function (PDF) may be more sensitive.   Here we investigate
the PDF for several cases. 

First consider a  scintillating point source with no TB averaging of $G$.
As is well known,
the intensity PDF is a 
one-sided exponential in the limit of no additive noise 
because
$G$ is a chi-square random variable (RV) with two degrees of freedom, 
$\chi^2_{2}$ (e.g. Goodman 1985).
TB averaging and extended sources
increase the number of degrees of freedom, and therefore
decrease $\misssq$,  as discussed in \S \ref{sec:ndof}.  
If TB averaging and source superposition are viewed 
as  combining statistically  independent RV,
$\Gbar$ is distributed as $\chi^2_{2\niss}$,  
\be
f_{\Gbar}(\Gbar) &\approx& 
    \frac{(\Gbar\niss)^{\niss}} {\Gbar\Gamma(\niss)} e^{-\Gbar\niss} U(\Gbar), 
\label{eq:pdfapprox}
\ee
where $\Gamma(x)$ is the gamma function and  $U(x)$ is the unit step function. 
In detail, however, the intensity (or visibility)
is the integral of variables that are statistically dependent, so
the $\chi^2_{2\niss}$ PDF is only an approximation  to the true PDF of G.    

The true PDF is calculated by solving a homogeneous Fredholm equation of
the second kind (Press \etal~ 1992, pp. 779-785)   
that results from expanding the propagator 
$g(\rvec, t, \rsvec, \nu)$  
onto an orthonormal set of eigenvectors 
$\psi_n$ (e.g. Goodman 1985, pp. 250-256)
  and requiring
that the expansion coefficients be statistically independent  
(a Karhunen-Lo\`eve expansion).   
The general case, where time-bandwidth averaging and source extension 
must be considered, requires solution of the eigenvalue problem
(see Appendix \ref{sec:appendixc})
\be
&~&
(TB)^{-1} 
\int_{t-T/2}^{t+T/2} dt^{\prime}
\int_{-B}^{+B} d\delta\nu\, \left ( 1 - \frac{\vert \delta\nu \vert}{B} \right)
\int d\rsvec_1\, \left [ I_s(\rsvec_1) I_s(\rsvec_2) \right]^{1/2}
\nonumber \\
&~&\quad\quad\quad\quad\quad \quad\quad\quad\quad\quad\quad 
\gamg(0, t^{\prime\prime} - t^{\prime}, \delta\nu, \rsvec_2-\rsvec_1)
\psi_n(t^{\prime}, \rsvec_1) 
=\lambda_n \psi_n(t^{\prime}, \rsvec_2),
\ee
where the $\lambda_n$ are the eigenvalues.
The time and frequency integrands are slightly different because the
wave propagator, $g$, is integrated over frequency before squaring
of the wavefield, while time-averaging occurs after squaring.

For the simple example of a time average of $G$ at discrete times
$t_j = j\Delta t, \, j = 1, \ldots, N$,
the eigenvectors $\psi_n(t_j)$ and eigenvalues $\lambda_n$ 
are solutions  of
\be
\sum_{i=1}^{N} \gamg(0, t_i-t_j,0,0) \psi_n(t_i) = \lambda_n \psi_n(t_j),
	\,\, n=1,\ldots,N.
\ee
The PDF of $\Gbar = N^{-1} \sum_i \vert g(t_i)\vert^2$ is the $N$-fold
convolution of one-sided exponential PDFs, each of which has
mean $\lambda_n/N$, because each coefficient $b_n$ in the expansion
of $g(t)$ is statistically independent and is a complex, Gaussian
RV.  The PDF can be written in the form
$f_G(G) = \sum_n c_n \exp(-G\,N/\lambda_n)$  where the $c_n$ are functions
of the eigenvalues,
$c_n = \lambda_n^{-1} \prod_{n^{\prime}\ne n} 
  (1-\lambda_{n^{\prime}} / \lambda_n)^{-1}$.
Frequency averaging behaves similarly.   
%An arbitrary source distribution yields the continuous  eigenvalue problem
%\be
%\int d\rsvec_1\, \left [ I_s(\rsvec_1) I_s(\rsvec_2) \right]^{1/2}
%\gamg(0, 0, 0, \rsvec_2-\rsvec_1)
%\psi_n(\rsvec_1) 
%= \lambda_n \psi_n(\rsvec_2).
%\ee

Goodman (1985) has shown that exact PDFs calculated in this way are
fairly well approximated by the $\chi^2_{2\niss}$ PDF with the appropriate
number of degrees of freedom given by Eq.~\ref{eq:ndof}.
The approximate PDF has the same mean and variance as the exact PDF.
In the limits $\niss\to 1$ and $\niss\gg 1$, the two PDFs become identical.  
Also, if an observable is calculated as the sum of strictly
independent DISS fluctuations with equal variances, then the 
exact PDF is the $\chi^2_{2\niss}$ PDF. 

\begin{figure}

\plotfiddle{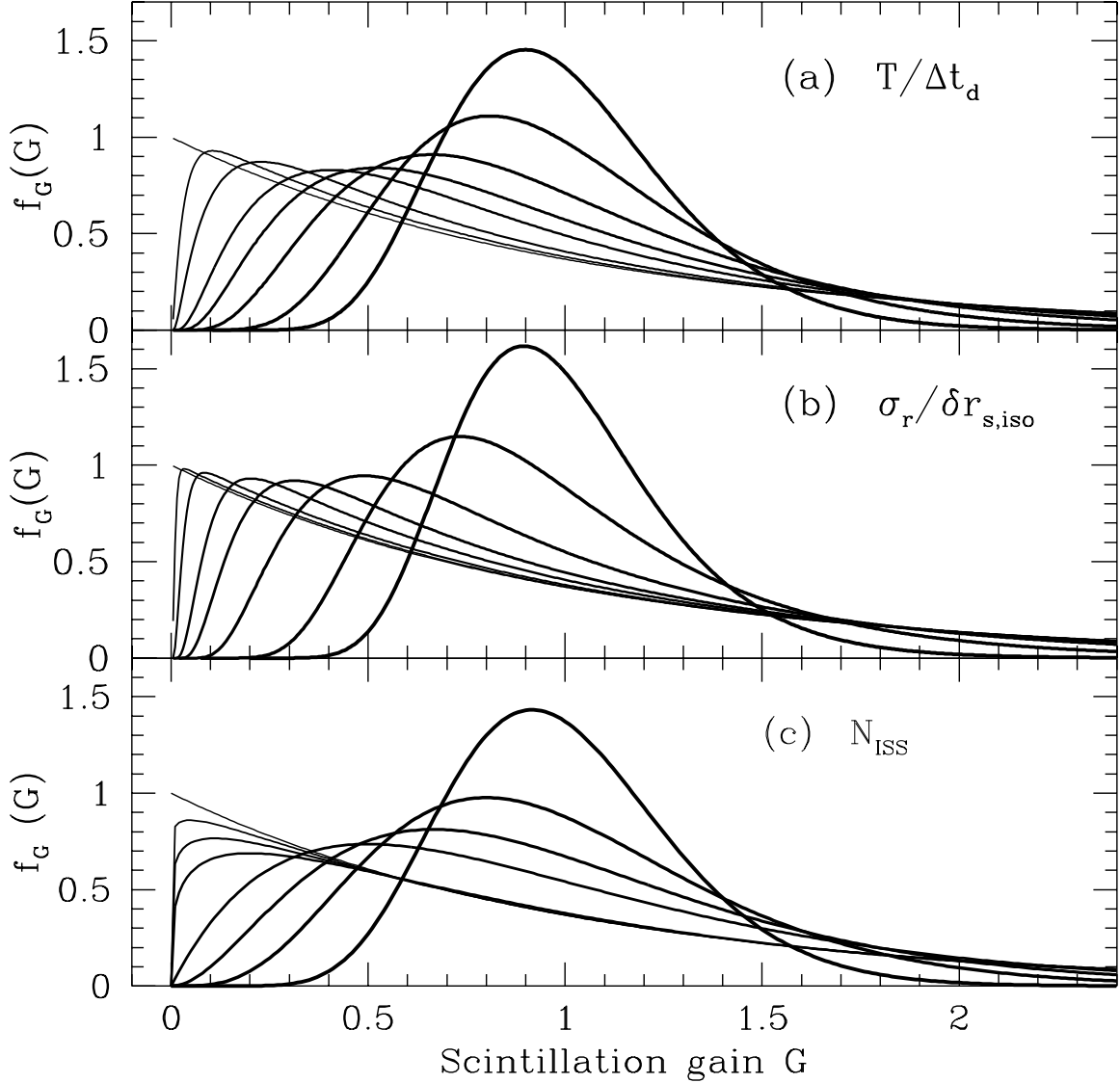}{6.0truein}{0}{80}{80}{-240}{-120}
\caption{
Probability density functions for the DISS
`gain' G for different amounts of time averaging and source extent. 
(a) PDFs vs. averaging time.
The different curves going from thinnest to thickest lines
are for $T/\dtd = $ 0.01, 0.5, 1, 2, 3, 5, 10 and 20.
The PDFs were determined by solving a  homogeneous Fredholm equation,   
as discussed in the text.
(b)
PDFs for 
Gaussian brightness distributions with
different sizes relative to the isoplanatic size.
The different curves going from thinnest to thickest lines
are for $\sigma_r / \drsiso = $ 0.01, 0.05, 0.1, 0.2, 0.3, 0.5, 1.0
and 2. These curves were determined by solving a two-dimensional Fredholm
equation.
(c) Approximate PDFs  given by a $\chi^2_{2\niss}$ PDF with  
number of degrees of freedom $2\niss$,  
where $\niss = 1, 1.05, 1.125, 1.25, 2, 3, 5$ and 12. 
}
\label{fig:fggtriptich}
\end{figure}

%Figure \ref{fig:eigen1} 
Figure \ref{fig:fggtriptich} shows exact PDFs 
for different instances of time averaging and finite source size.
The results are for
a Kolmogorov wavenumber spectrum ($\alpha = 5/3$) but do not differ
substantially for a square-law structure function.   
Figure \ref{fig:fggtriptich}a is a sequence of PDFs for time-averaging and
for a point source, while
Figure \ref{fig:fggtriptich}b is a sequence of PDFs for finite source
sizes but with no time-averaging. 
Also shown in Figure \ref{fig:fggtriptich}c 
 are approximate PDFs based on the $\chi^2$ PDF for different
numbers of degrees of freedom given by $2\niss$, where values of 
$\niss$ are chosen to yield PDFs of similar variance as in 
Figure \ref{fig:fggtriptich}a,b. 

%Figure \ref{fig:eigen2} shows PDFs for several source sizes when
%time-bandwidth averaging is negligible.   
Comparison of the panels in Figure  
\ref{fig:fggtriptich} indicates that time averaging
and source extent produce similar forms for the PDF of G.   
This conclusion verifies the notion that, from a statistical point of view,
TB-averaging and source extension produce like effects in scintillation
fluctuations.  The figure also supports the notion that one may use
the approximate  $\chi^2$ PDF to make calculations rather than
solving the Fredholm equation for every case.  This is useful because
in cases where time and frequency averaging as well as source extent
are important, the Fredholm solution may not be obtainable.

\subsection{PDF of Visibility Fluctuations}

\def\Ncal{{\cal N}}
\def\Ncaliave{{\langle {\cal N}_i\rangle}}

The PDF for $G$ derived in the previous section excludes contributions from
source fluctuations such as those that arise from the amplitude modulated
noise model.    
%For sources with arbitrary brightness distributions,
%it is difficult numerically  to include such fluctuations in the exact
%calculation  because four-dimensional eigenvectors must be considered. 
%Here we derive the PDF for the TB-averaged  visibility  that uses the
%approximate form for $f_G(G)$ based on the number of degrees of freedom
%associated with the finite source size.
Here we give a nearly exact treatment that takes into account
all fluctuations.
For simplicity, we assume that observation
 baselines do not resolve either the
source or the seeing disk from the scattering.  In this case, 
we write the average visibility across  baseline $\bvec_{ij}$ between
two sites $i$ and $j$ as
\be
\Gambar \approx  \Gbar\Iave + \Ncaliave\delta_{ij}  + X + C, 
\label{eq:GambarSAMN}
\ee
where 
$\langle I \rangle$ is the mean source intensity,
$\delta_{ij}$ is the Kronecker delta, 
$\langle \Ncal \rangle$ is the mean of 
$\Ncal_i\equiv\vert n_i \vert^2$ (the background noise intensity),
$X$ is a real Gaussian RV with zero mean, and  
$C$ is a complex Gaussian RV with zero mean.
Source fluctuations  are described by $X$ which includes noise fluctuations
associated with $m(\rsvec, t)$ and amplitude fluctuations associated with
$a(\rsvec, t)$ in the amplitude modulated noise model. 
$C$ includes additive 
radiometer noise combined with source noise fluctuations, but is uninfluenced
by source amplitude fluctuations.
Expressions for $\sigma_X^2$ and $\sigma_C^2$ are given 
in Appendix~\ref{sec:vispdfAMN}.

\def\inorm{{i}}

The PDF for the visibility magnitude is calculated 
by successively integrating over
the PDFs for the different, independent terms in Eq.~\ref{eq:GambarSAMN},
as done in Appendix ~\ref{sec:appendixc}.
The PDF for the scaled visibility magnitude, 
$\gamma =  \Gambarmag / \sigC$, is 
\be
\fgam(\gamma) &=& 
	\int d\Gbar\, f_{\Gbar}(\Gbar) 
	\int dX\, f_X(X)
	   \left [
		\gamma e^{-\half (\gamma^2 + \Gbar^2 \inorm^2) } I_0(\gamma \inorm) 
	   \right ]_{\inorm = (\Iave+ X/G)/\sigC}  
\label{eq:fgam11}
\ee
where $I_0$ is the modified Bessel function.  
The integrand factor in square brackets is the 
Rice-Nakagami PDF of  a signal phasor added to complex noise
(e.g. Thompson, Moran \& Swenson 1991, p. 260). 

\begin{figure}
\plotfiddle{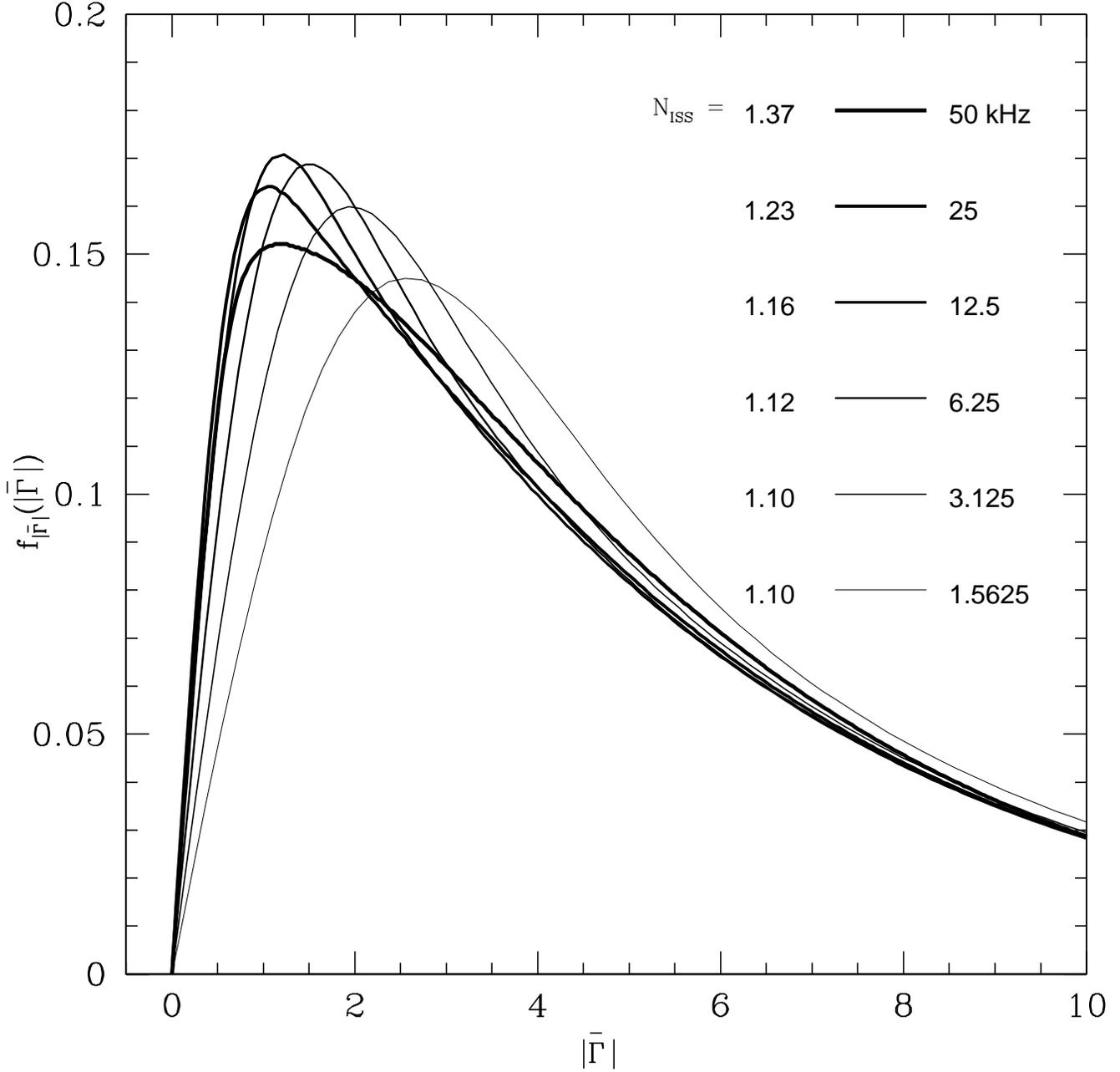}{6.0truein}{0}{90}{90}{-280}{-140}
\caption{
Probability density function for the visibility magnitude
for different bandwidths for  cases where 
%$\Aave / \Nave = 5$. 
$\Aave / \Nave = 0.14$.  
%$\npulses = 112$ and $T = 1.12$ ms. 
The different curves are for different bandwidths, as labelled,
which correspond to different numbers of ISS fluctuations, also as
labelled.   This figure corresponds to cases where ISS
variations and the complex term X are both included in 
Eq.~\ref{eq:fgam1}-\ref{eq:fgam}.
}
\label{fig:pdfvsbw-yesISS-yesX}
\end{figure}

To demonstrate the importance of various terms and factors, we
first show, in Figure \ref{fig:pdfvsbw-yesISS-yesX}, the visibility
PDF when we vary the bandwidth, taking into account that $\niss$
varies as we do so.   There is a tradeoff in discerning the underlying shape of
the DISS gain PDF (which encourages use of narrow bandwidths)     
and maximizing S/N, which favors larger bandwidths:
as we decrease the bandwidth there is less
averaging of the scintillations but contributions from noise (the $X$ and
$C$ terms) increase,  thus widening the PDF.  

\begin{figure}
\plotfiddle{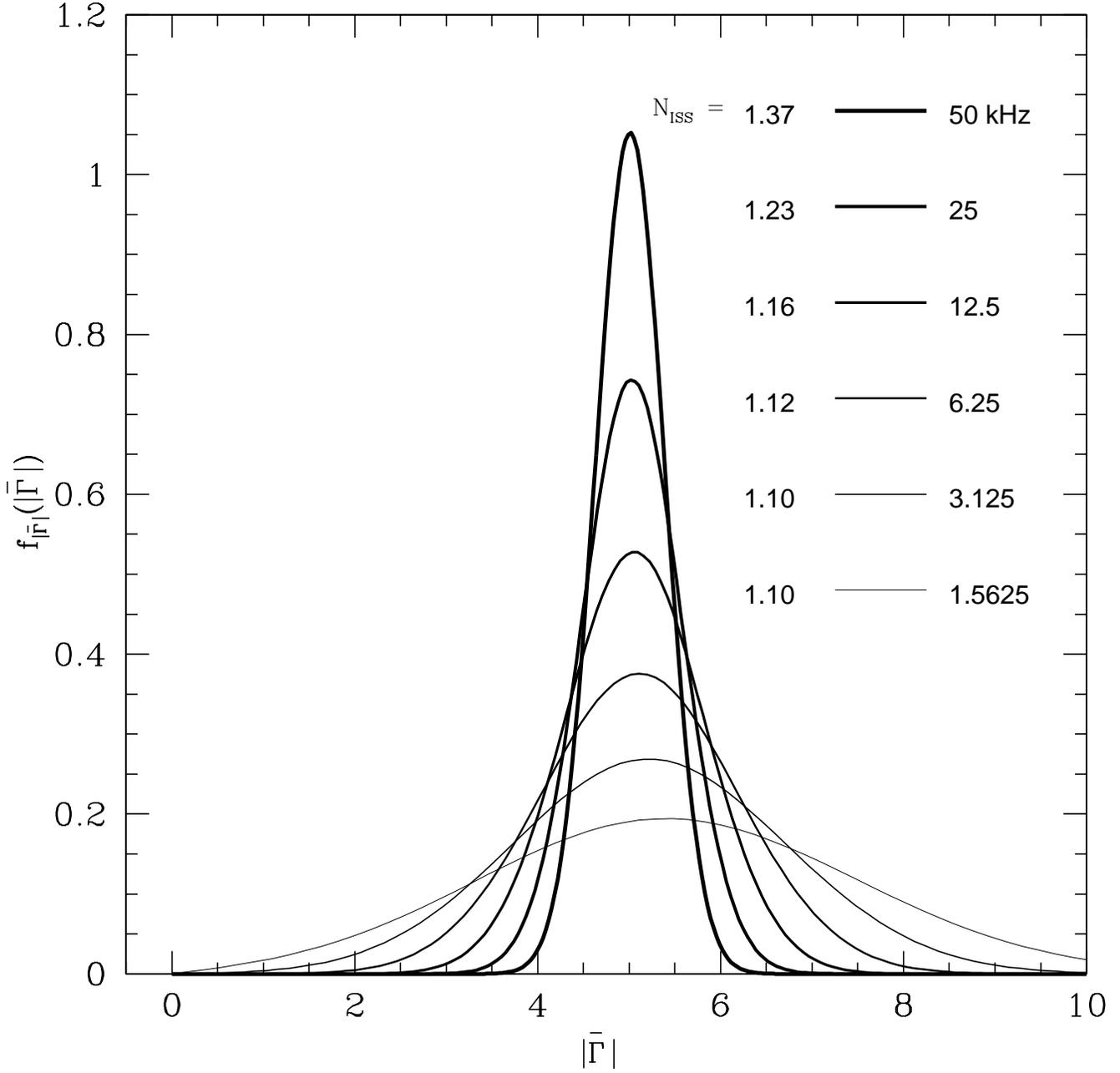}{6.0truein}{0}{90}{90}{-280}{-140}
\caption{
Same as Figure \ref{fig:pdfvsbw-yesISS-yesX} except
that ISS variations and the pulsar-noise term have been
turned off 
by making the PDF's for $G$ and $X$ delta functions, 
$\delta(G-1)$ and $\delta(X)$, in   
Eq.~\ref{eq:fgam2}-\ref{eq:fgam}.
}
\label{fig:pdfvsbw-noISS-noX}
\end{figure}

\begin{figure}
\plotfiddle{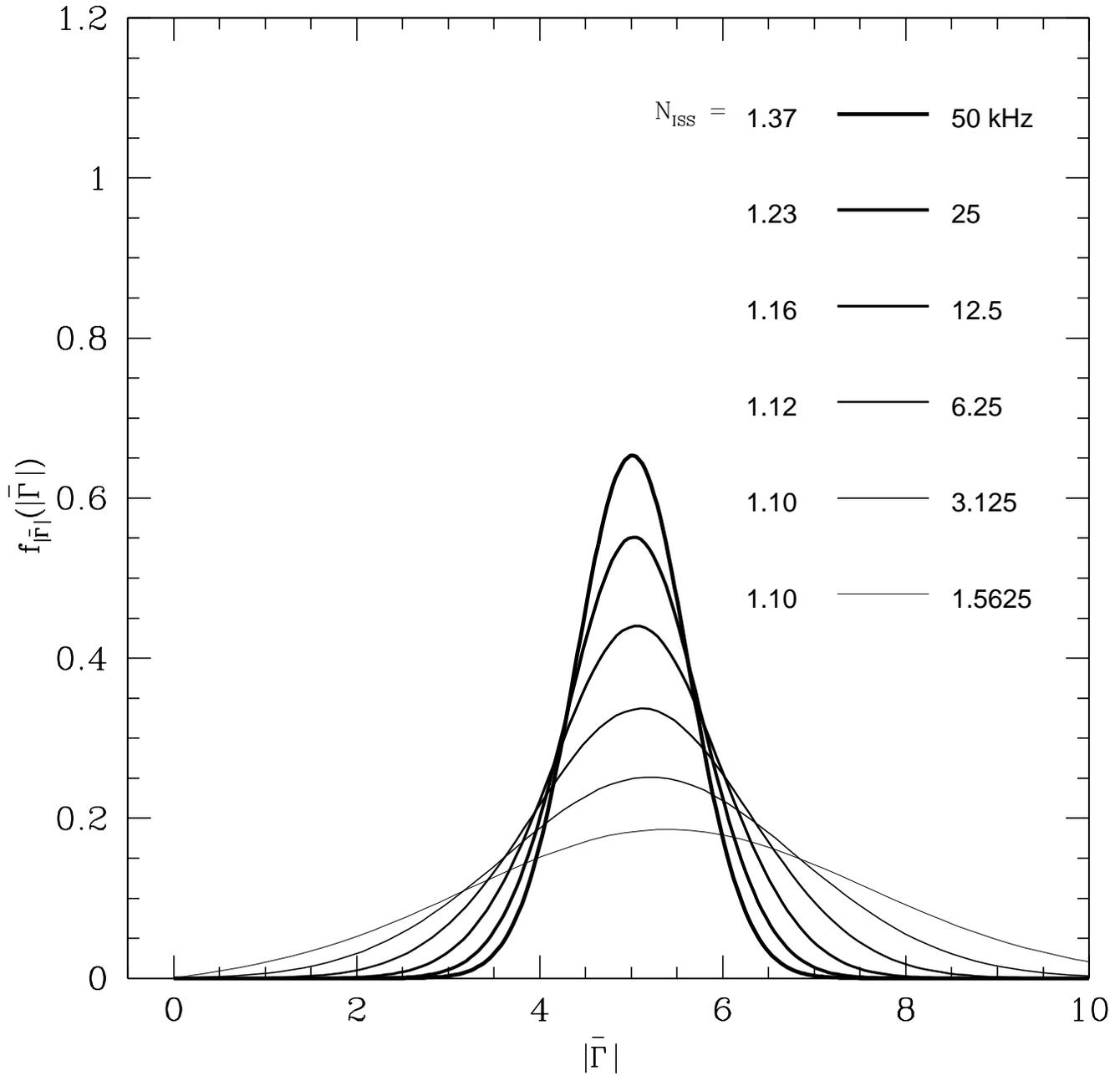}{6.0truein}{0}{90}{90}{-280}{-140}
\caption{
Same as Figures 
\ref{fig:pdfvsbw-yesISS-yesX}-\ref{fig:pdfvsbw-noISS-noX}
except that here ISS variations are turned off but the
pulsar noise variations $X$ are turned on.
}
\label{fig:pdfvsbw-noISS-yesX}
\end{figure}

\begin{figure}
\plotfiddle{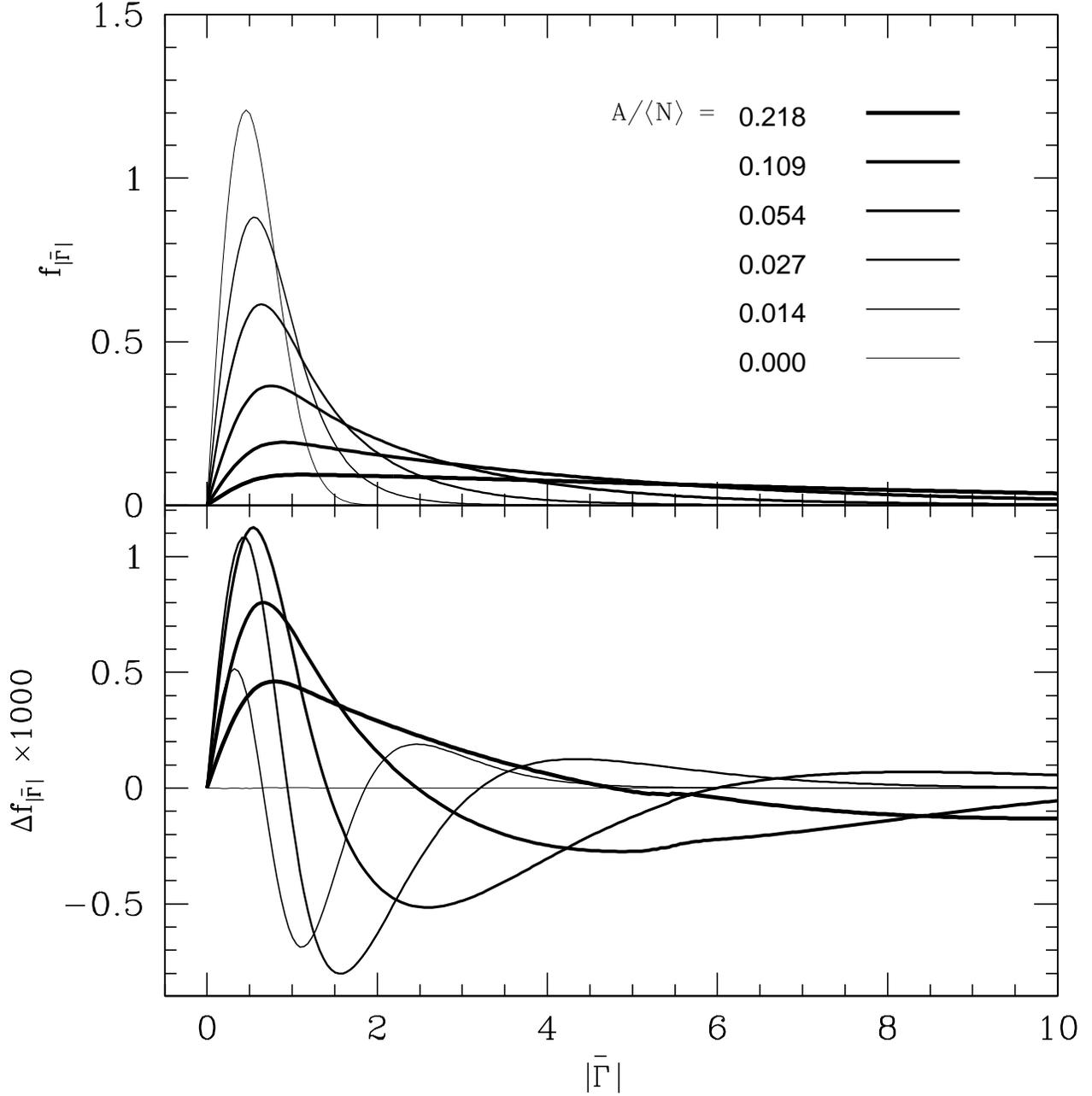}{6.0truein}{0}{90}{90}{-280}{-140}
\caption{
Top:  Visibility PDF for different source strengths. 
The ratio of signal strength to mean noise strength is
shown.   The PDF's were calculated for $\niss = 1.22$ and
a bandwidth of 25 kHz.
Bottom:  Difference between the true PDF and the PDF 
while ignoring pulsar fluctuations.  A positive value
means that the true PDF exceeds the latter PDF.
}
\label{fig:pdfvsa}
\end{figure}

We compare the $X$ and $C$ terms in 
Eq.~\ref{eq:GambarSAMN} with the $G\Iave$ term, which
dominates when the source is strong. 
First, we calculate the PDF for $\vert \Gambar \vert$ with
various terms excluded. 
Figure \ref{fig:pdfvsbw-noISS-noX} shows the visibility PDF when there
are no DISS and no  intrinsic fluctuations, just additive noise. 
Figure \ref{fig:pdfvsbw-noISS-yesX} shows the visibility PDF with the
intrinsic fluctuations turned on for a fairly high S/N observation,
but still with no DISS variations.   These curves indicate that 
source fluctuations can contribute significantly to the shape of the PDF.
%{\bf Show a family of curves for different S/N; perhaps normalize
%by something else besides $\sigma_+$.}
Figure \ref{fig:pdfvsa} shows the PDF for different source intensities
(top panel)
along with (bottom panel) the difference  between the true PDF
and the PDF where pulsar fluctuations are ignored.    The difference
vanishes when the source intensity is zero and $\sim 1$\%
for finite source intensities.   The error is largest near the peak
of the PDF. 

\begin{figure}
\plotfiddle{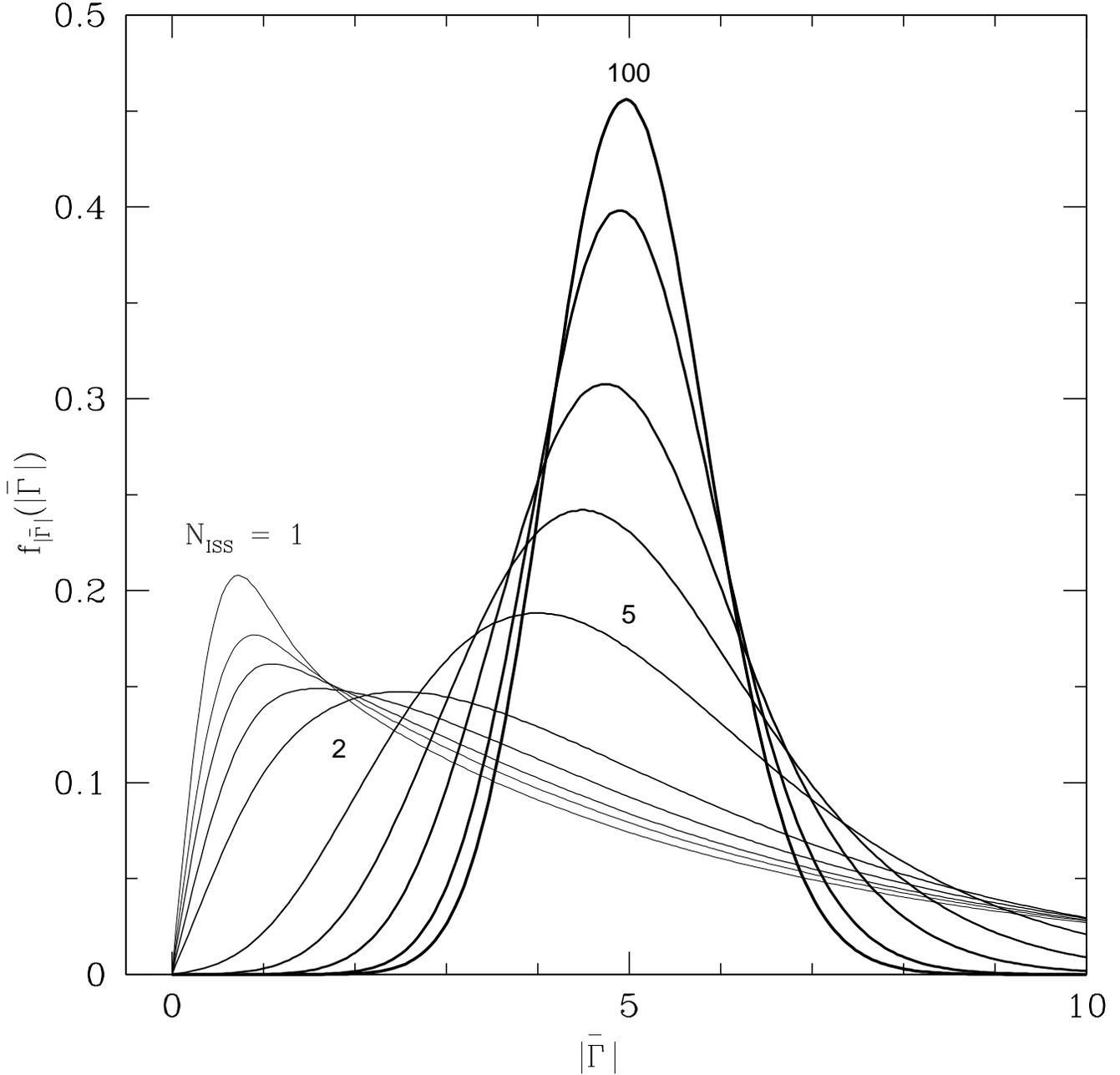}{6.0truein}{0}{90}{90}{-280}{-140}
\caption{
The PDF of the visibility magnitude 
when DISS is included with varying numbers of
degrees of freedom, $2\niss$, for 
$\niss = 1, 1.125,  1.25, 1.5, 2, 5, 10, 20, 50$ and 100,
with some values labelled.
Results are shown for 
$\langle A \rangle / \sqrt{N_i N_j} = 0.14$.
%Results are shown for
%$\eta_i = \eta_j = 7.8$.
% and $m_A = m_M = 1$.   
As $\niss \to \infty$, the PDF tends toward
a Gaussian function. 
}
\label{fig:figpdfsingle}
\end{figure}

Figure \ref{fig:figpdfsingle} 
shows the PDF when DISS is included with varying numbers of
degrees of freedom, $2\niss$, but with constant bandwidth.  This case would
apply to observations of sources with different intrinsic  sizes
or to a point source observed with varying amounts of time averaging. 
As $\niss \to \infty$, the PDF tends toward a Gaussian form. 
%Results are shown for
%$\eta_i = \eta_j = 7.8$, a signal-to-noise ratio appropriate for
%observations of the Vela pulsar, as discussed later. 
%We have scaled  $\vert \Gambar \vert$
%by $\sigplus \equiv \sigC(\Iave=0)$, the rms of the complex phasor
%$C$ when there is no signal.  
%The PDF of $\vert\Gambar\vert/\sigplus$ is then 
%$(\sigplus/\sigC) \fgam( \Gambar/\sigC)$.   
The results indicate that the net PDF for
$\Gambarmag$ is 
extremely sensitive to the number of degrees of freedom in the DISS. 
Also, it is evident that a pure point source can have statistics
that mimic those given by an extended source if there is sufficient
TB averaging. 

Comparison of Figures \ref{fig:figpdfsingle}
and \ref{fig:pdfvsbw-yesISS-yesX}  indicates that 
some of the changes in
PDF shape evident in Figure \ref{fig:figpdfsingle} that might be due
to, say, effects of source size,   are indeed masked by any changes in
resolution bandwidth in the signal processing. 

\subsubsection{When is Pulsar Noise Important?}

In \S\ref{sec:vispdfAMN} we give expressions for different terms in
the visibility variance, including those involving pulsar fluctuations
and one involving radiometer noise.  All fluctuations diminish, of course,
with increased averaging time.  However, the relative sizes of
the fluctuating terms are independent of the averaging time.  In studies
of the shape of the visibility or intensity PDF, pulsar fluctuations will   
be comparable to radiometer noise fluctuations when the pulsar signal 
strength satisfies
\be
G\Iave \gtrsim \left (\frac{\Niave\Njave} {B\,W_I} \right )^{1/2},
\ee
where $\Iave$ is the source flux density and $W_I$ is its characteristic
time scale (either intrinsic or imposed by some sampling scheme);
$B$ is the bandwidth.  For a pulsar, $W_I$ would be the sample window
in pulse phase and $\Iave$ the flux density in that window.

%\ noindent
%{\bf Refer to Gwinn results here.}
\subsection{Cross-Correlation Functions}

As described in Appendix~\ref{sec:appb-cross}, when the intensity is
measured for two sources and cross correlated, the total cross
modulation index includes only an ISS term because noise and source
fluctuations do not correlate.  Defining the cross-correlation of 
the average intensities as in Eq.~\ref{eq:CIbar} and defining the  cross
modulation index as
\be
\mcross12(\bvec,\taubar) = 
\frac{C_{\Ibar,12}(\bvec, \taubar) - I_1 I_2}{I_1I_2},
\ee
we find that $\mcross12$ is identically equal to
$\misssq$ of Eq.~\ref{eq:miss2}.     

The cross modulation index can be used in two ways to detect
source structure or spatial offsets between two sources.  First, if
its maximum  is smaller than unity, quenching of DISS by source structure
is signified.    Also, if the cross-modulation maximizes at a
time lag $\taubar\ne 0$ for zero baseline ($\bvec=0$), that signifies
a significant offset between sources.  We show these cases in the 
Appendix \ref{sec:appendixb} and will apply this method
to pulsar observations in a later paper.  

\section{Inferring Source Size from Intensity \& Visibility Statistics}
\label{sec:bayes}

\def\hatmiss2{{ {\hat m}_{\rm ISS}^2}}
\def\hatmgam2{{ {\hat m}_{\rm \Gamma}^2}}
\def\nisstotal{{N_{\rm ISS,TOTAL}}}
\def\nisstotalsqrt{{N^{-1/2}_{\rm ISS,TOTAL}}}

Given measurements of the visibility of intensity that have time and frequency
resolutions that resolve DISS, it is possible to place constraints on source
size by comparing visibility fluctuations with those expected from the 
time and frequency resolutions and as a function of source structure. 
Here we outline  general Bayesian procedures that first analyze 
visibility/intensity data and then are restricted to a simpler analysis of
just the modulation index.

\subsection{ Likelihood Analysis of Visibility \& Intensity Fluctuations}

Given a set of visibility (or intensity) measurements
\be
\left \{
  \Gambar_i(\bvec_{i}), i=1,N 
\right \}
\ee
we can normalize them by the rms off-source noise, $\sigC$, and use the
PDF of $\gamma \equiv \Gambar/\sigC$ (c.f. Eq.~\ref{eq:fgam11}) to calculate
a likelihood function
\be
{\cal L} = \prod_{i} \fgam(\gamma_i).  
\ee
The likelihood function depends on  numerous parameters that 
could be estimated by maximizing ${\cal L}$.   
These include parameters that describe the source,
$\Thetavec_S$, such as source structure and flux density;
those that describe wave propagation through the ISM,
$\Thetavec_{\bf ISM}$, including distances, type of medium (e.g. Kolmogorov),
and scintillation parameters ($\dnud, \dtd$);
and those that characterize the receiver and telescope system,
$\Thetavec_{\rm R}$, including the telescope gain and system temperature.
Many of these parameters will be known from auxiliary observations.  

Denoting all parameters collectively as 
$\Thetavec = 
   \left( \Thetavec_{\rm S}, \Thetavec_{\rm ISM}, \Thetavec_{\rm R} \right)$,
we identify the data probability 
$P({\cal D} \vert \Thvec)$ as the likelihood function
and we calculate, in standard Bayesian fashion
(e.g. Gregory \& Loredo 1992), the posterior PDF for the
parameters as
\be
P(\Thvec \vert {\cal D}) &=& 
	\frac   {
		  P(\Thvec) P({\cal D} \vert \Thvec)
		}
		{
		  \int d\Thvec\, P(\Thvec) P({\cal D} \vert \Thvec)
		}
= \frac	{
		  {\cal L}
	 	}
		{
		  \int d\Thvec \, {\cal L}
		},
\label{eq:Bayes}
\ee
where $P(\Thvec)$ is the prior PDF for the parameters and the denominator
normalizes the PDF.   The second equality follows if we assume that the 
parameters have a flat prior PDF.   In the case where, 
{\it a priori}, we know 
many of the parameters, we adopt delta function priors
and marginalize them by integrating over those parameters.  For example, if
we wish to derive the PDF of only source parameters, we would integrate over
$d\Thvec_{\rm ISM}\, d\Thvec_{\rm R}$ to obtain
$P(\Thvec_{\rm S}\vert {\cal D})$. 

To apply this approach we need to solve the multidimensional 
Fredholm problem that includes source extent and time-bandwidth averaging
if we want to use the exact PDF for scintillation gain.  Alternatively,
we could use the approximate PDF based on the $\chi^2$ PDF
(Eq.~\ref{eq:pdfapprox}) by calculating the effective number of degrees of 
freedom associated with source extent and TB averaging.   Another, simpler, 
approach is to analyze only the second moment of the visibility/intensity
fluctuations, as we now consider. 

\subsection{ Inferring Source Size from Modulation Indices}

The squared modulation index $\misssq$ can be calculated through
direct estimation of moments or by fitting a PDF shape to a histogram
of visibilities (or average intensities).  
Such estimates of  $\hatmiss2$ 
are typically made from  data that span a large
number of scintles, $\nisstotal\gg 1$.  
This number is approximately
\be
\nisstotal \approx 
\left ( 1 + \zeta \frac{T_{\rm tot}}{\dtd} \right )
\left ( 1 + \zeta \frac{B_{\rm tot}}{\dnud} \right ),
\label{eq:nissapprox}
\ee
where the total observing time and bandwidth may be written as
$T_{\rm tot} = N_T T$ and $B_{\rm tot} = N_{\nu} B$,
using $T$ and $B$ as the basic resolutions in time and frequency
defined previously.   The characteristic time and frequency scales
for DISS are $\dtd$ and $\dnud$, respectively.
The factor $\zeta \approx 0.2-0.3$ takes
into account that scintles are not packed tightly in the 
$\nu-t$ plane.  In Cordes (1986), I conservatively used
$\zeta = 0.1$ whereas a more accurate calculation yields
the values presented here. 

\def\siggam{{\sigma_{\Gamma^2}}}
\def\sigm2{{\sigma_{m^2}}}
The fractional estimation error on $\hatmiss2$ due
to the finite number of scintles is
$\sigm2\approx 2\misssq\nisstotalsqrt$.   
Invoking the Central Limit Theorem for $\nisstotal$, 
we expect $\hatmiss2$ to
have PDF, $N(\hatmiss2, \sigm2)$.

We write the likelihood function in terms of the PDF for 
$\hatmiss2$ estimated from data and using  
the model for $\misssq$
(as given by Eq.~\ref{eq:miss2}): 
\be
{\cal L} = \left ( 2\pi \siggam^2 \right )^{-1/2}
	\exp\left\{
		-\frac{1}{2\sigm2}
		 \left [ \misssq(\Thvec_{\rm S}) - \hatmiss2\right ]^2
	\right \},
\ee
where $\Thvec_{\rm S}$ is a vector of parameters that represents the 
source structure.
Bayes' theorem can then be applied according to Eq.~\ref{eq:Bayes} to 
derive the posterior PDF for $\Thvec_{\rm S}$.  
For the Gaussian brightness distribution of Eq~\ref{eq:gaussbright},
the posterior PDF is simply a one-dimensional PDF 
  $f_{\sigma_r}(\sigma_r)$
for the sole source parameter $\sigma_r$.

Uncertainties in the application of this inference scheme include
the systematic errors associated with not knowing the true form of the
structure function for the medium and also the statistical errors 
in the measured modulation index, $\hatmiss2$ and in the DISS parameters
$\dnud$ and $\dtd$.  We address these uncertainties in Paper II.

\def\RGeps{{R_{\Gamma_{\vareps}}}}

%\section{Future Directions}\label{sec:future}

\section{Summary and Conclusions}\label{sec:summary}

We have derived a general methodology for analyzing diffractive 
interstellar scintillation fluctuations that is applicable 
to single aperture and interferometric observations.    
In this paper, we considered only the strong scattering regime where
the scattered wavefield has Gaussian statistics. 
The method explicitly takes into account time-bandwidth averaging that is often
used in the statistical analysis of such observations.  Such averaging
modifies the statistics in a way that is identical to the effects
of extended source structure.   We show that time-bandwidth averaging
and extended source structure both increase the number of degrees
of freedom in the scintillations from the minimum value of two
that describes the fully modulated, Gaussian wavefield of the scintillations.

Our methodology can be applied to any radio source in the strong 
scattering regime,
including compact active galactic nuclei and gamma-ray burst afterglows.
In another paper, we will address sources of these types and we will also
consider scintillations in the weak and transition scattering regimes. 

In Paper II
we apply our results to the recent VLBI observations of the Vela
pulsar by Gwinn \etal~ (1997)
and find that the scintillation statistics may be accounted for fully
by time-bandwidth averaging.
  Any contribution from extended source
structure is less than an upper limit of about  400 km at the 95\% confidence
interval.  This upper limit on the  transverse extent is substantially larger      
than the size expected from conventional models that place radio
emission well within the light cylinder of the pulsar  and close to the
surface of the neutron star.

I thank Z. Arzoumanian, S. Chatterjee, C. R. Gwinn, H. Lambert, M. McLaughlin,  
and B. J. Rickett for useful discussions
and H. Lambert and B. J. Rickett for making available their 
numerically-derived autocovariance functions for Kolmogorov media. 
This research was supported by NSF grant 9819931 to Cornell
University and by NAIC, which is managed by Cornell University
under a cooperative agreement with the NSF.   

\clearpage
\appendix
\centerline{\bf APPENDICES}
\section{Scintillating Amplitude Modulated Noise Model}
\label{sec:appendixa}

Here we derive a general statistical model that incorporates incoherent
summing in the source and wave propagation throughout the interstellar medium.

The narrowband (scalar) electric field incident on an aperture and selected by
a feed antenna and by a bandpass receiver
may be written in the form
\be
E_{\Delta}(t) = Re \left \{ \vareps(t) \exp(-i\omega_0t)\right \},
\ee
where $\omega_0$ is the center frequency and the complex, baseband
wavefield is $\vareps$ (e.g. Thomas 1969).   
The baseband field is often explicitly extracted
through quadrature mixing schemes in heterodyned radio receivers
(e.g. Thompson, Moran \& Swenson 1991, p. 150).

Early work on pulsars 
modeled $\vareps(t)$ as amplitude modulated
noise (AMN) with additive background and receiver noise:
\be
\vareps(t) =  a(t) m(t) + n(t), 
\label{eq:amn}
\ee 
where $m$ and $n$ are complex, Gaussian wavefields that describe
intrinsic source noise and additive noise,
respectively.   The factor $a(t)$ is
a real modulation function that describes source variations 
on time scales much longer
than the reciprocal center frequency or reciprocal bandwidth;  
otherwise the statistics of $a(t)$ are arbitrary.
The additive noise has a modulation index
$m_N = [\langle \vert m \vert^4 \rangle / 
        \langle \vert m \vert^2 \rangle^2 -1 ]^{1/2} = 1.$  

The AMN model was
first presented by Rickett (1975), who attributed complex, Gaussian
statistics to $m(t)$.  For this case the modulation index is also
unity, $m_M = 1$.
Cordes (1976b) considered Poissonian shot-noise statistics for 
$m(t)$ based on physical models for pulsar emission;    
for Poissonian noise, $m_M \ge 1$.

Empirical tests on pulsars 
(Cordes 1976a; 
Hankins \& Boriakoff 1978;
Cordes \& Hankins 1979; 
Bartel \& Hankins 1982)
show consistency of $m(t)$ with Gaussian statistics on time scales as short
as $\sim 1 \,\,\mu s$.  Tests in the time domain resort to investigation of
the relative amplitudes of various terms in the autocorrelation function
of the intensity (Rickett 1975; Cordes 1976a; Bartel \& Hankins 1982).
Tests in the frequency domain, with the same conclusion,  
use the autocorrelation function of the spectrum (Cordes \& Hankins 1979).   
The model predicts
that there is frequency structure in the spectrum of a single pulse
with characteristic bandwidth equal to the reciprocal of the time duration
of $a(t)$.   This frequency structure averages out as multiple pulses are
summed.

Tests on OH and H$_2$O masers
(Evans \etal~ 1972; Moran  1981)
show that maser emission also conforms to the AMN picture.   
It is expected that {\it any} radio source can be described
by AMN because large numbers of particles contribute to the observed signals
and most, if not all, natural sources 
involve incoherent superposition of radiation from
incoherent or coherent emissions from individual radiators. 
Thus AMN should apply to gamma-ray burst sources and 
the most compact AGNs that show intra-day variability.

\subsection{Amplitude Modulated Noise for Extended Sources}

We model extended sources as follows.
First, the baseband field produced by a point source at
location $(\rsvec,z=0)$ is
\be
\vareps_s(\rsvec, t) = a(\rsvec, t)  m(\rsvec, t),
\label{eq:vareps}
\ee
where $a(\rsvec, t)$ is the (real) amplitude modulation and $m(\rsvec, t)$
is complex Gaussian noise (Rickett 1975). 
Here and everywhere, vectors are two dimensional and perpendicular
to the line of sight.
The quantity $\vareps_s$ is the field emitted per unit area at the
source, uninfluenced by propagation
(either through free space or a turbulent medium), except that 
we include the dependence on distance from the source on 
its mean amplitude, for simplicity.
For a steady source, $a(\rsvec, t)$, is constant
in time.  For pulsars, it describes the periodic envelope of pulses
that modulates the underlying noise process, $m(\rsvec, t)$. 
The corresponding mean intensity is, using $A\equiv a^2$, 
\be
I_s(\rsvec, t)  
	= \langle \vert \vareps_s(\rsvec, t) \vert^2 \rangle 
	= \langle A(\rsvec, t)\rangle.
\ee
In most of this paper we assume stationary statistics,
so $I_s(\rsvec, t) \to I_s(\rsvec)$.

The total measured field is the integral over source components
\be
\vareps(\rvec, t) = \int d\rsvec\, \vareps_s(\rsvec, t)
			g(\rvec, t, \nu, \rsvec),
\label{eq:epsmodel}
\ee
where we include a multiplicative
 propagation factor, $g(\rvec, t, \nu, \rsvec)$, defined  in the next
section.

The noise, with stationary statistics, has  a correlation function
\def\Gammtwo{{\Gamma_{2m}}}
\def\Gammtwoc{{\Gamma^*_{2m}}}
\def\Gammfour{{\Gamma_{4m}}}
\be
\Gammtwo(\rsvec_1, \rsvec_2, t_1,  t_2) = 
	\langle m(\rsvec_1, t_1) m^*(\rsvec_2, t_2) \rangle = 
    	\delta(\rsvec_1-\rsvec_2) \Delta(t_2 - t_1),
\ee
where the asterisk denotes conjugation and $\Delta(\tau)$ is an Hermitian 
function having unit amplitude, $\Delta(0)=1$,
and width approximately equal to the inverse of the receiver bandwidth. 
Angular brackets denote an ensemble average,
except where noted.
The noise fourth moment is  the standard dual sum of products
for a complex Gaussian process, 
\be
\Gammfour(\rsvec_1, \rsvec_2, \rsvec_3, \rsvec_4, t_1, t_2, t_3, t_4) 
	&=& 
	\langle 
		m(\rsvec_1, t_1) m^*(\rsvec_2, t_2) m(\rsvec_3, t_3) m^*(\rsvec_4, t_4) 
	\rangle \nonumber\\
	&=& 
	\Gammtwo(\rsvec_1, \rsvec_2, t_1,  t_2)  
	\Gammtwo(\rsvec_3, \rsvec_4, t_3,  t_4) \nonumber \\ 
	&+&
	\Gammtwo(\rsvec_1, \rsvec_4, t_1,  t_4) 
	\Gammtwoc(\rsvec_2, \rsvec_3, t_2,  t_3).  
\label{eq:Gammfour}
\ee

\subsection{Propagation Through a Thin Diffracting Screen}

\def\rprime{{rvec^{\prime}}}
\def\rvecp{{\rvec^{\prime}}}

Consider the following geometry: 
a point source at $(\rsvec, 0)$,
a thin screen at $(\rvecp, D_s)$ and 
an observer at $(\rvec, d = \ds+D)$. 
The screen changes only the phase of incident
waves.
Under the narrowband approximation ($\Dnu\ll\nu$) 
(so that all phase factors may be considered constant over the band)
%and assuming further
%that the screen phase $\phi(\rvecp, t)$  varies slowly and hence very little
%over the time $\Dnu^{-1}$, 
the propagated baseband field for a point source
at $\rsvec$ is (e.g. Goodman 1985)
\be
\vareps(\rvec, t, \rsvec) = 
	(i\lambda\Dbar)^{-1}
	\int d\rvecp \, 
		e^{i\phi(\rvecp, t - c^{-1}{\cal D}_{23})}
		e^{ik {\cal D}_{13}}
		\vareps_s(\rsvec, t - c^{-1}{\cal D}_{13}), 
\ee
where
$\Dbar \equiv (\ds^{-1} + \dso^{-1})^{-1}$ and
the integral is normalized so that a screen with zero phase
yields simply a delayed version of the emitted field,
$\vareps_s(\rsvec, t-c^{-1}d)$.
Under the paraxial approximation (transverse scales  much smaller
than line-of-sight distances),
\be
{\cal D}_{13} &=& {\cal D}_{12} + {\cal D}_{23} \nonumber \\
{\cal D}_{12} &\approx& \ds + 
		\displaystyle \frac{\vert \rvecp - \rsvec \vert^2}{2\ds}
		\nonumber \\ 
{\cal D}_{23} &\approx& \dso + 
		\displaystyle \frac{\vert \rvec - \rvecp \vert^2}{2\dso}.
		\nonumber  
\ee
We assume further that variations in  propagation times, 
$c^{-1} {\cal D}_{13}$ and $c^{-1} {\cal D}_{23}$ 
(as a function of relevant source locations $\rsvec$ and 
screen exit points $\rvecp$) 
are negligible compared to the characteristic
variation time scales for $\phi(\rvecp, t)$ and $\vareps(\rsvec, t)$. 
%Assuming that the screen phase much more slowly than does the signal,
%which varies on a time scale $\sim $(\Delta\nu)^{-1}$, we require
$c^{-1}\Delta\nu({\cal D}_{13}-d) \ll 1$. 
Typically, 
$\vareps_s$ varies on time scales
of order the reciprocal bandwidth (e.g. 100 $\mu s$ or less) while
$g$ varies, due to the changing geometry, 
on time scales of seconds to hours or more, 
for the situations we wish to consider.
Therefore, for $\vareps_s$ to be factored out of the integral, we 
require
$c^{-1}\Delta\nu({\cal D}_{13}-d) \ll 1$. 
This simply means that any time smearing from differential arrival times
must be less than the time resolution of the signal.
Though we assume in the remainder that the inequality  
is satisfied, we point out that there are many instances where
it is not, corresponding to the well known `pulse broadening'
effect (e.g. Rickett 1990).   As a rule of thumb, when pulse broadening
is important, scintillations are difficult to resolve in time and frequency.
And when scintillations are important, the pulse broadening can be too
small to be important, as we assume here.

We can now write the propagated field for a single point source as
\be
\vareps(\rvec, t, \rsvec) = 
    \vareps_s(\rsvec, t-d/c) g(\rvec, t-D/c, \nu, \rsvec),
\label{eq:sigmoda}
\ee 
where $g$ is the propagator, 
\be
g(\rvec, t, \nu, \rsvec) = (i\lambda\Dbar)^{-1}
	\int d\rvecp \, 
		\exp\left\{ 
		i\left[\frac{k}{2}
		\left ( 
		\ds^{-1} \vert \rvecp - \rsvec \vert^2  + 	
		\dso^{-1} \vert \rvec - \rvecp \vert^2 	
		\right)
		+ \phi(\rvecp,t) \right]
		\right\},
\ee
and we use $k=2\pi c^{-1}\nu$. 
The normalization of $g$ yields it to be simply a unit modulus
phase factor when the screen phase is zero and also it yields
$\langle \vert g \vert^2\rangle = 1$ when the screen phase has
Gaussian statistics. 
%We have also shifted the time origin of $\phi$ by $D/c$.
In the following we will ignore the delays $d/c$ and $D/c$ in 
Eq.~\ref{eq:sigmoda} in our notation. 
The time dependence of $g(\rvec, t, \nu, \rsvec)$ arises from motions of
source, observer and medium, which influence all terms in the exponent in
the integrand. 
Absent any random phase screen (i.e. $\phi = 0$),
$g$ is simply a complex phase factor that describes free-space
propagation,
\be
g(\rvec, t, \nu, \rsvec) = e^{ik\vert \rsvec - \rvec \vert^2/2d}.
\ee

\subsection{Propagator Second Moment for a Thin Screen}

\def\Rvec{{\bf R}}

The propagator's second moment across a baseline $\bvec$,
at two times separated by $\tau$ for two point sources
at $\rsvec_{1,2}$ and at two frequencies separated by $\delta\nu$ is
% note commented lines below with the \frac{Dnu} factor are
% neglibible under the paraxial approximation and in strong scattering
% where relevant transverse length scales are much smaller than the Fresnel scale.
\be
\Gamg(\bvec, \tau, \delta\nu, \delta\rsvec) &=& 
	\langle 
		g(\rvec, t, \nu, \rsvec)
		g^*(\rvec+\bvec, t+\tau, \nu+\delta\nu, \rsvec+\delta\rsvec)
	\rangle \nonumber \\ 
	&=& e^{-i\psi}  \gamg(\bvec, \tau, \delta\nu, \delta\rsvec) 
\label{eq:Gamg-a}\\
\psi    &=&kd^{-1} \{
             		(\rsvec-\rvec)\cdot \Delta \Rvec
 	    		+ \half \vert \Delta\Rvec\vert^2  
%    			+ \half \left(\frac{\Dnu}{\nu}\right) 
%			\vert \Delta\Rvec + \rsvec - \rvec \vert^2 
		   \} \\
\Delta \Rvec &=& [\delta\rsvec-\bvec+(\vpvec-\vobsvec)\tau]. 
\label{eq:Gamg-b}
\ee
%The phase $\psi$ is accurate to second order in all quantities.
There is no term $\propto \Dnu$ in $\psi$ because it is negligible
according to the narrowband assumption made earlier, i.e. that
$c^{-1}\Delta\nu({\cal D}_{13}-d) \ll 1$. 

The real, second moment $\gamg$ for zero frequency lag is
\be
\gamg(\bvec, \tau, \delta\nu=0, \delta\rsvec) &=& 
	e^{-\half \dphi(\bvec, \tau,  \drsvec)} 
 	\label{eq:gamg-a} 
\ee
where the phase structure function $\dphi$ and its arguments are given by
\be
\dphi(\bvec, \tau, \drsvec) &\equiv& 
	\left\langle
	[ \phi(\rvec,t) - \phi(\rvec+\beffvec, t) ]^2
	\right\rangle = \left(\frac{\beff}{\bd}\right)^{\alpha} \\ 
\beffvec &=& (\ds/d) \bvec + \veffvec\tau + (D/d)\drsvec \\
\veffvec &=& (\ds/d)\vobsvec + (1-\ds/d)\vpvec 
    - \vismvec \label{eq:veffvec-appendixa}
%\rsvec &=& \half (\rsvec_1 + \rsvec_2) \\
%\drsvec &=& \rsvec_2 - \rsvec_1.
\ee
The scaling exponent  for the structure function, $\dphi$,
is $\alpha$, which takes on a value
$\alpha=5/3$ for a Kolmogorov spectrum in some instances.
The $e^{-1}$ scale of $\vert\gamg\vert^2$ is $b_e$, 
$\vpvec$ is the pulsar velocity, $\vobsvec$ is the observer's velocity
and $\vismvec$ is the velocity of the scattering material in the ISM.
The normalized autocovariance function for 
$G = \vert g \vert^2$, which we call the intensity ``gain,'' 
is (in the strong scattering, or Rayleigh, regime)
\be
\gamG (\bvec, \tau, \delta\nu, \drsvec) \equiv 
\langle G(\rvec, t, \nu, \rsvec_1) \, G(\rvec+\bvec,t+\tau,\nu+\delta\nu,\rsvec_2) 
-1
= \vert \gamg(\bvec, \tau, \delta\nu, \drsvec) \vert^2,
\label{eq:gamG-a}
\ee
and the mean intensity is
\be
I(\rvec, t) 
%	= \Gamma_s({\bf 0}, t) 
	= \int d\rsvec\, I_s(\rsvec, t) 
	= \int d\rsvec\, \langle A(\rsvec, t) \rangle. 
\ee
Note that $\langle G \rangle = 1$.
We emphasize that the form for $\gamg$ in the second equalities
of Eq.~\ref{eq:Gamg-a} and Eq.~\ref{eq:gamG-a} relies 
on the assumption that the scattered wavefield
is sufficiently randomized that Gaussian statistics apply.

For nonzero frequency lags, $\gamg$ generally must be obtained through
appropriate numerical integration (e.g. Lambert \& Rickett 1999;
Lee \& Jokipii 1975).     For the special case of a square-law
structure function, a closed form expression is available for a thin screen
(Chashei \& Shishov 1976; Cordes \etal~ 1986; Gupta \etal~ 1994).

\subsection{Extension to An Arbitrarily Thick Medium}

The results presented so far were derived explicitly for a thin screen.
For a medium in which scattering occurs with variable
strength all along the line of sight, the results may be extrapolated 
quite simply and generally.  Assuming that the net measured field is 
still Gaussian, it is a matter of simple 
geometry to work out the form of the equivalent phase structure function.
Following Lotova \& Chashei (1981) and  Cordes \& Rickett (1998) we
use the same results as in Eq.~\ref{eq:Gamg-a}-\ref{eq:visfn-b}
but make the replacements,
\be
\dphi(\bvec, \tau, \drsvec) &=& (\lambda r_e)^2 f_{\alpha}
	\int_0^d ds\, \cnsq(s) \vert \beffvec(s) \vert^{\alpha} \\ 
\beffvec(s)  &=& (s/d) \bvec + \veffvec(s)\tau + (1-s/d) \drsvec \\
\veffvec(s) &=& (s/d)\vobsvec + (1-s/d)\vpvec - \vismvec(s),  
\ee
where 
\be
f_{\alpha} = \frac{8 \pi^2}{\alpha 2^{\alpha}} 
\frac{\Gamma (1-\alpha/2)}{\Gamma (1+\alpha /2)} .
\ee
and 
$\cnsq$ is the coefficient in the wavenumber spectrum for electron
density variations (Cordes \& Lazio 1991; Armstrong, Rickett \& Spangler 1995).
For the case $\cnsq(s) \propto \delta(s-\ds)$, 
we retrieve the thin-screen results of the previous section.

\subsection{Visibility Function}

The ensemble mean visibility function is
the product of the true source visibility and the
propagator's second moment (ignoring a phase factor) 
\be
\langle \Gameps(\bvec, t, \taui) \rangle  = 	
\langle \vareps(\rvec, t) \vareps^*(\rvec+\bvec, t+\taui) = 
			\Delta(\taui) 
			\gamg(\bvec, 0, 0, 0)
			\Gams(\bvec, t),
\label{eq:visfn-a}
\ee
where we have designated the interferometer lag as $\taui$.  This must
match any geometrical time delays to within the reciprocal of the 
receiver bandwidth.
The source visibility is the usual Fourier transform of the 
brightness distribution,
\be
\Gams(\bvec, t) = 
	\int d\rsvec \, e^{+ik\dpo^{-1}\rsvec\cdot\bvec}
	I_s(\rsvec, t).
\label{eq:visfn-b}
\ee
%We maintain  time dependence in the visibility function for cases
%where the source intensity is intrinsically  time dependent.  
%This is the case for pulsars on short time scales, and masers and 
%active galactic nuclei on much longer times.
The time lag $\taui$ in Eq.~\ref{eq:visfn-a} is zero to within 
a very small time (of order the reciprocal bandwidth; see Thompson,
Moran \& Swenson 1991).  By contrast,  the time lag in Eq.~\ref{eq:Gamg-a}  
extends over very long times, seconds to hours, that characterize
DISS fluctuations.

\clearpage
\section{Intensity \& Visibility Fluctuations}
\label{sec:appendixb}
%\input appendixb.tex

% appendix on intensity and visibility fluctuations

%\subsection{Fourth Moments}

\def\RGeps{{R_{\Gamma_{\vareps}}}}
\def\Rfoureps{{R_{4\vareps}}}
\def\RI{{\rm R_I}}

Here we consider variations in the time-averaged intensity and visibility
for the model of Appendix~\ref{sec:appendixa}.    We derive the
modulation fractions of these quantities taking into account scintillations
and intrinsic and additive noise.   Our expressions will include
any averaging of the scintillation modulation over frequency as well as time. 

We define the general fourth moment for the narrowband field
$\vareps$ ({\it sans} additive noise)
\be
\Rfoureps(\bvec, t_1, t_2, t_3, t_4) &=& 
	\langle 
	\vareps(\rvec, t_1)	
	\vareps^*(\rvec, t_2)	
	\vareps(\rvec+\bvec, t_3)	
	\vareps^*(\rvec+\bvec, t_4)	
	\rangle,
\label{eq:fourepsdef}
\ee
where $\vareps(\rvec, t)$ is given by Eq.~\ref{eq:epsmodel}.
Expanding out, $\Rfoureps$ involves fourth-order moments of
$a(\rsvec, t)$ and $m(\rsvec, t)$ from the AMN model of 
Appendix~\ref{sec:appendixa} and of the propagator $g(\rvec, t, \nu, \rsvec)$. 
We use the fact that $a, m$, and $g$ are statistically independent.
Also, $m$ and $g$ are complex gaussian processes so their fourth moments
are dual sums of products of their second moments, as in Eq.~\ref{eq:Gammfour}.

The fourth moment of of $a(\rsvec, t)$ ends up as the second moment of its
square, $A \equiv a^2$, which we assume has the form   
\be
\langle A(\rsvec_1, t_1) A(\rsvec_2, t_2) \rangle 
	= \langle A(\rsvec_1, t_1) \rangle \langle  A(\rsvec_2, t_2) \rangle 
	  [ 1 + m_A^2 \rho_A(\rsvec_2 - \rsvec_1, t_2 - t_1)].
\label{eq:rhoAdef}
\ee
This form assumes stationary statistics for $A$ with a correlation function
$\rho_A$ and  modulation index $m_A$.
As we show in paper II, assuming stationary statistics is 
not  restrictive, even for pulsars which are highly nonstationary across
pulse phase but appear to have
stationary statistics when a fixed pulse phase is considered.
For other radio sources with intrinsic variations much longer than those
of pulsars, we may consider $\langle A(\rsvec, t)\rangle$ to be constant
in time, with $m_A = 0$, at least over a typical observation time of
minutes.

Our treatment also includes any variations of the scintillation propagator, $g$,
across the  bandwidth of the narrowband signal, $\vareps(\rvec, t)$.   
By partitioning $\vareps$ into subbands in which the propagator is 
piecewise constant and between which the emitted signal $\vareps_s$ is 
statistically independent, we easily can incorporate finite bandpass
effects while being consistent with our earlier assumption about
the narrowband signal.\footnote{The net effect is that intensities from the
subbands add while the intensity autocorrelation function discussed later 
involves an average over frequency lag.}

\subsection{Autocorrelation Functions of Time Average Quantities}

The time-averaged intensity,
\be
\Ibar(\rvec, t) = T^{-1} \int_{t-T/2}^{t+T/2} dt^{\prime}\, 
	I(\rvec, t^{\prime}),
\ee
has autocorrelation function
\be
R_{\Ibar}(\bvec, \tau) \equiv  
	\langle \Ibar(\rvec, t) \Ibar(\rvec+\bvec, t+\tau) \rangle 
	&=& T^{-2} \int\int_{t-T/2}^{t+T/2} 
		dt_a\, dt_b \, 
		\RI(\bvec, t_b - t_a+\tau)\\
        &=& T^{-1} \int_{-T}^{+T} d\tau^{\prime} 
			\left( 1 - \frac{\vert \tau^{\prime} \vert}{T} \right)
			\RI(\bvec, \tau^{\prime}+\tau).
\label{eq:RIbar}
\ee
The integrand is given by 
\be
\RI(\bvec,\tau) = \Rfoureps(\bvec, t_1, t_1, t_1+\tau, t_1 + \tau).  
\ee

Similarly, we consider the visibility,
\be
\Gameps(\bvec, t, \taui) = \vareps(\rvec, t)\vareps^*(\rvec+\bvec, t+\taui)
\ee
and its time average,
\be
\Gambar(\bvec, t, \taui) = T^{-1} \int_{t-T/2}^{t+T/2} dt^{\prime}\, 
			\Gambar(\bvec, t^{\prime}, \taui).
\ee
The autocorrelation, analogous to Eq.~\ref{eq:RIbar}, is
\be
R_{\Gambar}(\bvec, \taubar; \taui) 
        = T^{-1} \int_{-T}^{+T} d\tau^{\prime} 
			\left( 1 - \frac{\vert \tau^{\prime} \vert}{T} \right)
			R_{\Gambar}(\bvec, \tau^{\prime}+\taubar;\taui),
\label{eq:RGambar}
\ee
and involves the integrand
\be
R_{\Gambar}(\bvec,\taubar; \tau) = 
 	\Rfoureps(\bvec, t_1, t_1+\taubar, t_1+\taubar+\tau, t_1+\tau).  
\ee
Note that we distinguish here between the lag associated with the
definition of the visibility, $\taui$, and the lag $\taubar$ with which we
consider the autocorrelation of the visibility. 

\subsection{Modulation Indices}

We are most interested in the normalized variances of the time-average
intensity  and visibility.  These are defined in terms of the 
autocorrelation functions  as
\be
m^2_{\Ibar}(\bvec, \tau) &\equiv&
	\frac{\displaystyle{ R_{\Ibar}(\bvec, \tau) - \Iave^2}}{\Iave^2} \\  
m^2_{\Gambar}(\bvec, \tau) &\equiv&
	\frac{\displaystyle{ R_{\Gambar}(\bvec, \tau) - 
		\vert\Gamepsave\vert^2}}{\Iave^2},  
\ee 
where we normalize by the mean intensity in both cases.

\def\mpsrasq{{m^2_{\rm PSR}}}
\def\mpsrmsq{{m^2_{\rm NOISE}}}

The total modulation index squared for the intensity or visibility
is the sum of  three main terms:
\be
m^2(\bvec, \taubar) 	= \misssq(\bvec, \taubar) 
			+ \mpsrasq(\bvec, \taubar),
			+ \mpsrmsq(\bvec, \taubar),
\ee
where $\misssq$ measures the contribution from scintillations only,
$\mpsrasq$ measures the contribution from source amplitude fluctuations 
combined with scintillations, and 
$\mpsrmsq$ measures source noise fluctuations.  
Later (\S\ref{sec:additive})  we will also consider
the effects of additive radiometer fluctuations.    
For pulsars, $\mpsrasq$ includes pulse shape
variations and noise fluctuations.  For sources that are steady over an
observation span of minutes to hours (or more), $\mpsrasq = 0$. 
The `noise' term, $\mpsrmsq$, depends on source structure and corresponds
to the output of an intensity interferometer that is proportional to 
 the square of the 
visibility function (e.g. Hanbury-Brown 1974, pp. 48-49).

We can write these terms as
\be
\misssq(\bvec, \taubar) &=& 
	\Iave^{-2} \int\int d\xvec\, d\yvec \,
		\Irsvec(\xvec) \Irsvec(\yvec)
		\Qiss(\bvec, \taubar, \yvec - \xvec, T, B) \\
\mpsrasq(\bvec, \taubar) &=& 
	\Iave^{-2} \int\int d\xvec\, d\yvec \,
		\Irsvec(\xvec) \Irsvec(\yvec)
		\Qpsr(\bvec, \taubar, \yvec - \xvec, T, B) \\
\mpsrmsq(\bvec, \taubar) &=& 
	\Iave^{-2} \int\int d\xvec\, d\yvec \,
		\Irsvec(\xvec) \Irsvec(\yvec)
		\Qnoise(\bvec, \taubar, \yvec - \xvec, T, B) 
\ee

%\be
%\misssq(\bvec, \taubar) &=& 
	%\Iave^{-2} \int\int d\rsvec_1 d\rsvec_2 \,
		%%\langle A(\rsvec_1) \rangle
		%%\langle A(\rsvec_2) \rangle
		%\Irsvec(\rsvec_1) \Irsvec(\rsvec_2)
		%\Qiss(\bvec, \taubar, \rsvec_2 - \rsvec_1, T, B) \\
%\mpsrasq(\bvec, \taubar) &=& 
	%\Iave^{-2} \int\int d\rsvec_1 d\rsvec_2 \,
		%%\langle A(\rsvec_1) \rangle
		%%\langle A(\rsvec_2) \rangle
		%\Irsvec(\rsvec_1) \Irsvec(\rsvec_2)
		%\Qpsr(\bvec, \taubar, \rsvec_2 - \rsvec_1, T, B) \\
%\mpsrmsq(\bvec, \taubar) &=& 
	%\Iave^{-2} \int\int d\rsvec_1 d\rsvec_2 \,
		%%\langle A(\rsvec_1) \rangle
		%%\langle A(\rsvec_2) \rangle
		%\Irsvec(\rsvec_1) \Irsvec(\rsvec_2)
		%\Qnoise(\bvec, \taubar, \rsvec_2 - \rsvec_1, T, B). 
%\ee

The `$Q$' functions are defined in terms of integrals 
over time lag, like that in Eq.~\ref{eq:RIbar}  and over similar 
frequency-lag integrals, that we denote as
\be
\left \langle X(y) \right \rangle_{y,Y}
	\equiv Y^{-1} \int_{-Y}^{+Y} dy\, 
		\left( 1 -\frac{\vert y \vert }{Y}\right) X(y). 
\ee

\subsection{Intensity Fluctuations}

For intensity fluctuations, the Q functions are
\be
\Qiss(\bvec, \taubar, \delta\rsvec, T, B) 
	&=& \nonumber \\ 
&~&\!\!\!\!\!\!\!\!\!\!\!\!\!\!\!\!\!\!\!\!\!\!\!\!\!\!\!\!\!\!\!\!\!\!
\!\!\!\!\!\!\!\!\!\!\!\!\!\!\!\!\!\!\!\!\!
		\left \langle 
	   	  \gamG(\bvec, \tau^{\prime}+\taubar, \delta\nu, \delta\rsvec) 
		\right \rangle_{\taup, T; ~\delta\nu, B},
\label{eq:Qissdef-intensity} \\
\Qpsr(\bvec, \taubar, \delta\rsvec, T, B) 
	&=& \\
&~&\!\!\!\!\!\!\!\!\!\!\!\!\!\!\!\!\!\!\!\!\!\!\!\!\!\!\!\!\!\!\!\!\!\!
\!\!\!\!\!\!\!\!\!\!\!\!\!\!\!\!\!\!\!\!\!
     	m_A^2\, 
		\left \{
		\left\langle
			\rho_A(\delta\rsvec,\tau^{\prime}+\taubar) 
		\right\rangle_{\taup, T}
		+
		\left\langle
		  \rho_A(\delta\rsvec, \tau^{\prime}+\taubar) 
	   	  \gamG(\bvec, \tau^{\prime}+\taubar, \delta\nu, \delta\rsvec) 
		\right \rangle_{\taup, T; ~\delta\nu, B}
		\right.  
		\nonumber\\  
&~&\!\!\!\!\!\!\!\!\!\!\!\!\!\!\!\!\!\!\!\!\!\!\!\!\!\!\!\!\!\!\!\!\!\!
\!\!\!\!\!\!\!\!\!\!\!\!\!\!\!\!\!\!\!\!\!
		\left.
		+ R_{\Delta}(\taubar, T) e^{-ikd^{-1} \bvec\cdot\delta\rsvec}
		\rho_A(\delta\rsvec, 0)
 		\left[
	   	  \gamG(\bvec, 0, 0, 0) +  
		\left\langle
	   	        \gamG(0, 0, \delta\nu, \delta\rsvec) 
		\right\rangle_{\delta\nu, B}
		\right]
		\right \}, \nonumber \\ 
\label{eq:Qpsr-intensity}
\Qnoise(\bvec, \taubar, \delta\rsvec, T, B)&=& 
		R_{\Delta}(\taubar, T) e^{-ikd^{-1} \bvec\cdot\delta\rsvec}
 		\left[
	   	  \gamG(\bvec, 0, 0, 0) +  
		\left\langle
	   	        \gamG(0, 0, \delta\nu, \delta\rsvec) 
		\right\rangle_{\delta\nu, B}
		\right].  \,\,\,\,\, 
\label{eq:Qnoise-intensity}
\ee
We have made use of the lag-integrated noise correlation
\be
R_{\Delta}(\tau, T) \equiv T^{-1} \int_{-T}^{T} d\taup\,
	\left( 1 - \frac{\taup}{T} \right)
	\vert \Delta(\taup)\vert^2.
\ee
Recall that $\Delta(\tau)$ is a function with unit maximum amplitude 
[$\Delta(0) = 1$] and width of order the reciprocal bandwidth,
$B^{-1}$.  For integration times $T\gg B^{-1}$, we have
\be 
R_{\Delta}(\tau, T) \approx \frac{\wdelta}{T} U(T-\tau)U(T+\tau),
\ee  
where $U(x)$ is the unit step function and
$\wdelta \equiv \int d\tau\, \vert \Delta(\tau)\vert^2 \approx B^{-1}$
is the characteristic time scale of the noise fluctuations.

Note that for $T\gg\wdelta$, the terms involving $R_{\Delta}$ 
in $\Qpsr$ and $\Qnoise$ are much smaller than $\Qiss$. 
Also, for radio sources other than pulsars, $\Qpsr \equiv 0$ because $m_A = 0$. 

\subsection{Intensity Interferometry}

We can relate our results to those of Hanbury-Brown and Twiss
(e.g. Hanbury Brown 1974), who used the intensity autocorrelation function
to determine the magnitude of the source visibility function for optical
stars.   The
term of interest in their work corresponds to our $\Qnoise$, in particular
the second term which involves  
$
		\left\langle
	   	        \gamG(0, 0, \delta\nu, \delta\rsvec) 
		\right\rangle_{\delta\nu, B}.
$
The first term with $\gamG(\bvec, 0, 0, 0)$ vanishes for baselines much
larger than the Fried scale.  Also, the apertures used by Hanbury Brown 
and Twiss were larger than the Fried scale, causing aperture averaging
that we have not treated but which is analogous to time-bandwidth 
averaging.
For ground-based optical observations of stars, Hanbury-Brown and Twiss
used a bandwidth such that scintillations were constant over the band and
the stars they observed were much smaller than the isoplanatic scale.
In this case, $ \gamG(0, 0, \delta\nu, \delta\rsvec) \to 1$ and the 
effect was maximized.  Note also, however, that the amplitude of the effect
scales with $R_{\Delta}(\taubar, T)\approx \wdelta/T \approx (BT)^{-1}$, 
which is small for significant
time-bandwidth averaging.

\subsection{Visibility Fluctuations}

For the visibility we have,
using $\DV \equiv \vpvec - \vobsvec$, 
\be
\Qiss(\bvec, \taubar, \delta\rsvec, T, B) 
	&=& 
		\vert \Delta(\taui)\vert^2
		\left \langle 
	   	  \gamG(0, \tau^{\prime}+\taubar, \delta\nu, \delta\rsvec) 
	 	  e^{-ikd^{-1} \bvec\cdot[\delta\rsvec + \DV(\taup+\taubar)]}
		\right \rangle_{\taup, T; ~\delta\nu, B},
\label{eq:Qissdef-visibility} \\
\Qpsr(\bvec, \taubar, \delta\rsvec, T, B) 
	&=& \\
&~&\!\!\!\!\!\!\!\!\!\!\!\!\!\!\!\!\!\!\!\!\!\!\!\!\!\!\!\!\!\!\!\!\!\!
\!\!\!\!\!\!\!\!\!\!\!\!\!\!\!\!\!\!\!\!\!
     	m_A^2\, 
		\left \{
		  \vert \Delta(\taui)\vert^2
	 	  e^{-ikd^{-1} \bvec\cdot\delta\rsvec}
		\left\langle
			\rho_A(\delta\rsvec,\tau^{\prime}+\tau) 
	   	   	\gamG(\bvec, 0, 0, 0) 
		\right\rangle_{\taup, T} 
		\right.
		\nonumber \\
&~&\!\!\!\!\!\!\!\!\!\!\!\!\!\!\!\!\!\!\!\!\!\!\!\!\!\!\!\!\!\!\!\!\!\!
\!\!\!\!\!\!\!\!\!\!\!\!\!\!\!\!\!\!\!\!\!
		\left.
		+
		\left\langle
		  \rho_A(\delta\rsvec, \tau^{\prime}+\tau) 
	   	  \gamG(0, \tau^{\prime}+\taubar, \delta\nu, \delta\rsvec) 
	 	  e^{-ikd^{-1} \bvec\cdot[\DV(\taup+\taubar)]}
		\right \rangle_{\taup, T; ~\delta\nu, B}
		\right.  
		\nonumber\\  
&~&\!\!\!\!\!\!\!\!\!\!\!\!\!\!\!\!\!\!\!\!\!\!\!\!\!\!\!\!\!\!\!\!\!\!
\!\!\!\!\!\!\!\!\!\!\!\!\!\!\!\!\!\!\!\!\!
		\left.
		+ R_{\Delta}(\taubar, T) 
		\rho(\delta\rsvec, 0)
 		\left[
			1 + 	   	  
		\left\langle
	   	        \gamG(\bvec, \taui,  \delta\nu, \delta\rsvec) 
		\right\rangle_{\delta\nu, B}
		\right]
		\right \}. \nonumber \\ 
\label{eq:Qpsr-visibility}
\Qnoise(\bvec, \taubar, \delta\rsvec, T, B)&=& 
		R_{\Delta}(\tau, T) 
 		\left[
			1 + 	   	  
		\left\langle
	   	        \gamG(\bvec, \taui,  \delta\nu, \delta\rsvec) 
		\right\rangle_{\delta\nu, B}
		\right].  \,\,\,\,\, 
\label{eq:Qnoise-visibility}
\ee

\subsection{Effects of Additive Noise}
\label{sec:additive}

\def\mradsq{{m^2_{rad}}}
Results given so far for intensity and visibility fluctuations have considered
only the signal emitted by the source.   Including the additive radiometer noise 
$n$ as in Eq.~\ref{eq:amn}, we obtain an additional contribution to the total
modulation index, that we denote $\mradsq$:
For the time-average intensity and visibility, the 
contribution is
\be
\mradsq(\bvec_{ij}, \taubar) = 
\left \{
\begin{array}{ll}
		2\Iave^{-1}
		\Niave
		\delta_{ij} R_{\Delta}(\taubar, T)
		\left( 1 + \half \Iave^{-1} \Niave\right) 
		& \mbox{{\rm \, intensity fluctuations}}\\
\\
		\Iave^{-1}
		\Niave
		R_{\Delta}(\taubar, T)
		\left( 1 + \frac{\Njave}{\Niave} + \Iave^{-1} \Njave \right) 
		& \mbox{{\rm \, visibility fluctuations}}\\
		\end{array}
		\right.
\ee
We have labelled the baseline with $ij$ indices to represent the
$i$th and $j$th sites.
The Kronecker delta indicates that the contribution for the intensity
 holds only for single-site measurements for which $i$ and $j$ are equal.

\subsection{Cross Correlations of Time-Average Intensities}
\label{sec:appb-cross}

In some circumstances, we are interested in the cross correlation function of
the intensity between two sources that may or may not be scintillating
together.   Pulsars, for example, have
 different pulse components that  may come from spatially
different emission regions and it is possible to record or
compute intensities for
each component separately and cross-correlate them. 
We define the cross correlation as
\be
C_{\Ibar,12}(\bvec, \taubar) \equiv  
	\langle \Ibar_1(\rvec, t) \Ibar_2(\rvec+\bvec, t+\tau) \rangle. 
\label{eq:CIbar}
\ee
The utility of the cross correlation is that it is affected by both the 
separations of the two sources and the size of each source.   For example,
the time lag at which the CCF maximizes is determined by the separation of
the sources and by the effective velocity, $\veffvec$
(Eq.~\ref{eq:veffvec},\ref{eq:veffvec-appendixa}).

The cross correlation simplifies greatly if we assume that the amplitude-modulated
noise in each source is statistically independent from the other.   
For pulsars, this is a reasonable assumption in many cases, though in others
where there are drifting subpulse fluctuations that appear successively in
different pulse components, this is an approximation. 
Letting $I_{1,2}$ be the (ensemble) mean intensity of each source,
we find that 
\be
C_{\Ibar_{12}}(\bvec, \taubar) &\equiv&  
I_1(t) I_2(t+\taubar) \\
	&+&
	\int\int d\rsvec_1 d\rsvec_2
	\langle A(\rsvec_1, t) \rangle
	\langle A(\rsvec_2, t+\taubar) \rangle
	\langle 
		\gamma_G(\bvec, \taup+\taubar, \delta\nu, \rsvec_2 - \rsvec_1)
	\rangle_{\taup, T; \delta\nu, B}. \nonumber
\ee

For two point sources, one at $\rsvec_1$, another at $\rsvec_2$, that have
stationary statistics, the normalized crosscovariance is
\be
\gamma_{\,\Ibar_{12}} (\bvec, \taubar) \equiv
	\displaystyle\frac
		{C_{\Ibar_{12}}(\bvec, \taubar) - I_1 I_2}
		{I_1 I_2} 
		=
		\left\langle
		\gamma_G(\bvec, \taup+\taubar, \delta\nu, \rsvec_2 - \rsvec_1)
		\right\rangle_{\taup, T; \delta\nu, B}.
\ee

To illustrate the utility of the crosscovariance consider the case where
$\gamma_G$ is constant over the averaging intervals $T$ and $B$.
The time lag that maximizes $\gamma_{\,\Ibar_{12}}$ is the solution of
\be
	        \frac{\partial}{\partial\taubar}
	   	\dphi(\bvec, \taup+\taubar, \rsvec_2-\rsvec_1)
		= 0.
\ee
For a thin screen this becomes
(c.f. Eq.~\ref{eq:gamg}-\ref{eq:gamG}) 
\be
\frac{\partial}{\partial\taubar}  
\left \vert \left(\frac{\ds}{d}\right) \bvec + 
	\veffvec\taubar + \left(\frac{D}{d}\right)\drsvec \right\vert = 0, 
\ee
which has the solution
\be
\taumax	= -\frac
		{
		\veffvec\cdot [ (\ds/d)\bvec + (D/d)\drsvec ]
		}
		{
		\veff^2
		}.
\ee
The ability to estimate $\taumax$ with precision depends on its
value relative to the characteristic DISS time, which is the
width of $\gamG$ as a function of $\taubar$. Defining 
$\dtd$ using $\gamG(0,\dtd,0, 0) = e^{-1}$, we find that
$\dtd = \ld / \veff$, where $\ld$ is the characteristic 
diffraction scale, yielding
\be
\frac{\taumax}{\dtd} = -\frac  
		{
		\veffvec\cdot [ (\ds/d)\bvec + (D/d)\drsvec ]
		}
		{
		\ld \veff
		}.
\ee
For $\bvec = 0$,  we expect to identify $\drsvec \ne 0$
only if $(D/d)\drsvec$ is a sizable fraction of
 $\ld$ and also if the effective velocity
is not orthogonal to $\drsvec$. 
The definition of `sizable' depends on the number of
independent ISS fluctuations used in any estimate of the cross correlation
function, which is $\niss$ given by Eq.~\ref{eq:ndof}.   The error on
$\taumax$ $\sim \dtd\niss^{-1/2}$, so a three-sigma measurement requires
$\delta\rsvec_{\parallel} \gtrsim 3 \niss^{-1/2}(d\ld/D)$, where
$\rsvec_{\parallel} \equiv \rsvec\cdot\veffvec$.

Smirnova, Shishov \& Malofeev (1996) give a similar expression for
$\taumax$ that is based on a medium with a square-law structure function. 
Our result is more general.

\clearpage
\section{Probability Densities for Strong Scattering}
\label{sec:appendixc}

\def\inorm{{i}}

Here we derive probability density functions for the DISS gain in
strong scattering.   We use an exact treatment based on  Karhunen-Lo\`eve
expansions that take into account source extent and time-bandwidth
averaging.   Then we derive the PDF for the visibility and intensity
that takes into account all features of the amplitude modulated noise
model of Appendix \ref{sec:appendixa}.

\subsection{Exact Solution for the PDF of the Scintillation Gain}

The time average intensity may be written as
\be
\Ibar(\rvec, t) = T^{-1} \int_{t\pm T/2} dt^{\prime}
	\int d\rsvec\, I_s(\rsvec) \, 
	\vert \gbar(r, t^{\prime}, \nu, \rsvec) \vert^2
\label{eq:klsetup}
\ee
for a source with arbitrary brightness distribution $I_s$
modulated by DISS.   The DISS modulation $\gbar$ has been
bandwidth averaged in accord with the considerations of 
Appendix \ref{sec:appendixa},
\be
\gbar(\rvec, t, \nu, \rsvec) = 
	B^{-1} \int_{\nu\pm B/2} d\nu^{\prime}\,
	g(\rvec, t, \nu^{\prime}, \rsvec).
\ee
Following the approach described by Goodman (1985; pp. 250-252),
we expand $I_s^{1/2}(\rsvec) \gbar(\rvec, t, \nu, \rsvec)$ onto a
set of orthonormal basis vectors $\psi_n(t,\rsvec)$ with coefficients
$b_n$.  The orthonormality condition is
\be
T^{-1} \int_{t\pm T/2} dt\, \int d\rsvec\,
	\psi_n(t,\rsvec) \psi^*_{n^{\prime}}(t,\rsvec) = \delta_{nn^{\prime}}
\ee
and the $b_n$ are given by
\be
b_n = T^{-1} \int_{t\pm T/2} dt\, \int d\rsvec\,
	I^{1/2}_s(\rsvec) \gbar(\rvec, t, \nu, \rsvec)
	\psi_n(t,\rsvec). 
\label{eq:bn}
\ee
By requiring that 
$\langle b_n b^*_{n^{\prime}}\rangle = 
\langle \vert b_n \vert^2 \rangle \delta_{nn^{\prime}}$
(i.e. that the $b_n$ are statistically independent), the following eigenvalue
problem results:
\be
&~&
(TB)^{-1} 
\int_{t-T/2}^{t+T/2} dt^{\prime}
\int_{-B}^{+B} d\delta\nu\, \left ( 1 - \frac{\vert \delta\nu \vert}{B} \right)
\int d\rsvec_1\, \left [ I_s(\rsvec_1) I_s(\rsvec_2) \right]^{1/2}
\nonumber \\
&~&\quad\quad\quad\quad\quad \quad\quad\quad\quad\quad\quad 
\gamg(0, t^{\prime\prime} - t^{\prime}, \delta\nu, \rsvec_2-\rsvec_1)
\psi_n(t^{\prime}, \rsvec_1) 
=\lambda_n \psi_n(t^{\prime}, \rsvec_2),
\ee
where $\lambda_n = \langle \vert b_n \vert^2\rangle$ are the eigenvalues.
The time and frequency averaging are handled differently because the
wave propagator, $g$, is integrated over frequency before squaring
of the wavefield, while time-averaging occurs after squaring.

The expansion implies that
\be
\Ibar(\rvec, t) = \sum_n \vert b_n \vert^2
\label{eq:Ibarsum}
\ee
and that
\be
\langle I(\rvec, t) \rangle = \sum_n \lambda_n.
\ee
The expansion coefficients are gaussian distributed because the integral
Eq.~\ref{eq:bn} is a sum of gaussian variables.    
Therefore, each term in Eq.~\ref{eq:Ibarsum}, $\vert b_n \vert^2$, 
is exponentially distributed
and the intensity PDF is the convolution of each of these exponentials.

The convolution can be calculated through Fourier transforms and inverted
using the residue theorem yielding, for nondegenerate eigenvalues,
\be
f_I(I) = \sum_{n=1}^{N} c_n \, e^{-I/\lambda_n} U(I),
\ee
where $U(I)$ is the unit step function and the coefficients are given
by 
\be 
c_n = {\lambda_n}^{-1} {\prod_{n^{\prime}\ne n}^N}
	\left ( 1 - \lambda_{n^{\prime}}/\lambda_n \right )^{-1}.
\ee
When only one eigenvalue is important, as it is for a point source with
negligible time-bandwidth averaging, the PDF for I contains only a single
term with mean $\lambda = \Iave$.

\subsection{Visibility PDF for the AMN Model}
\label{sec:vispdfAMN}

Here we present an alternative derivation of the visibility PDF that
takes into account 
all source, propagation and additive-noise fluctuations.

The visibility function is the 
product of the narrowband fields from two sites 
(i and j)\footnote{We take the product at the same time in order 
to keep notation
simple.  In practice, a delay must be introduced to account for the
different optical path lengths to the two sites.   Our notation assumes this
has already been removed.},
\be
\Gamma(t) = \vareps_i(t) \vareps_j^*(t),
\ee
where $\vareps_{i,j} = g_{i,j} a\, m + n_{i,j}$ for
a point source that produces an identical field at the two sites.  The
propagator and the additive noise are both different at the two sites, 
in general. 

The instantaneous value of the visibility is
\be
\Gamma = \gigj A \, M + \ninj + a(g_i m n_j^* + g_j^* m^* n_i). 
\ee
We have simplified the notation, using $i,j$ to label spatial location
rather than using location and baseline vectors as we have used in previous
sections.
The first term is the scintillated pulsar signal, the second is due to additive
noise at the two sites, while the third term represents cross products.
If there were no scintillations, the visibility would be a noisy phasor
(from the pulsar) combined with complex noise.   Scintillations modify
the source phasor to make it complex, in general.  However, the pulsar
noise (from $A$ and $M$) are in phase with respect to the scintillations.

The (ensemble-average) mean visibility is
\be
\langle \Gamma \rangle  = 
\langle g_i g_j^* \rangle 
\langle A \rangle 
+
\langle N_i \rangle \delta_{ij}.
\ee

In practice, a time average is used to approximate the ensemble average,
with attendant errors.  We use the following notation for  
the time-averaged visibility: 
\be
\Gambar = \langle \Gamma(t) \rangle_{BT} \equiv
     {T}^{-1} 
	\int_{t-T/2}^{t+T/2}\, 
	\Gamma(t),
\label{eq:gamave}
\ee
where the subscript `BT' on the angular brackets 
denotes time averaging of a bandlimited process  
with bandwidth B.

By expanding $A$ and $M$ into mean values and
zero-mean fluctuations, e.g.  
$A = \langle A \rangle + \delta A$ and 
$M = \langle M \rangle +\delta M$, 
we can write the time-average visibility  as
\be
\Gambar = \gigjave_{BT} \Aave \Mave + \Niave
	\delta_{ij} + X + C,
\label{eq:Gambar}
\ee 
where $\delta_{ij}$ is the Kronecker delta.
The first term is due to the source, the second term is the mean
system noise for a single site observation ($i=j$), and the last two terms
are fluctuations,
\be
X &=& \langle 
	\gigj \left[\delta A + \delta M\Aave + \delta A\delta M \right ]
      \rangle_{BT} \nonumber \\
C &=& \langle a(g_i m n_j^* + g_j^* m^* n_i) \rangle_{BT} 
        + \ninjave_{BT}  - \Niave\delta_{ij}.
\nonumber 
\ee
We separate $X$ and $C$ because in useful limiting cases, discussed in the
next subsections, 
they become, respectively, real and complex processes.   
For $i=j$, $C$ also becomes real.
Moreover, $X$ is in phase with the phasor term,
$\gigjave_{BT} \Aave \Mave$,
while $C$ is randomly phased.

If there are no intrinsic fluctuations, $X$  vanishes 
and $C$ then depends only on additive noise.  
However, the AMN model demands that there be source fluctuations
even if there are no amplitude modulations.
We let $\Mave\equiv \langle \vert m \vert^2\rangle = 1$ 
without any loss of generality.

Though we have assumed a point source to arrive at Eq.~\ref{eq:Gambar},
the equation also applies to extended sources that are spatially incoherent.
Spatial incoherence yields summation of contributions from different
source elements that also imply Gaussian statistics.
%The ensemble average of Eq.~\ref{eq:Gambar} is 
%\be
%\langle \Gambar \rangle = \Gamg(\bvecij, 0) \Aave + \Niave\delta_{ij}
%\label{eq:Gambarave}
%\ee

All terms in $X$ and $C$  are uncorrelated,  so variances of individual
terms sum to yield the total variance.   
We assume time-bandwidth averaging such that 
$BT\gg 1$;  thus
$X$ and $C$ become Gaussian random variables (GRVs) by the
Central Limit Theorem.  However, we assume  $BT$  is small enough so that 
the DISS factor, $\gigjave_{BT}$, is not a GRV. 
For now, holding $g_i(t)$ and $g_j(t)$ as fixed realizations of the DISS
fluctuation (i.e. not averaging over an ensemble for these quantities),
we find that $X$ has 
PDF \footnote{$N(0,\sigX^2)$
denotes a Gaussian PDF of a real variable  with zero mean and variance
$\sigX^2$ while 
$N_c(0, \sigC^2)$ denotes a complex Gaussian quantity having 
real and imaginary parts with equal variances, $\sigC^2$.},  
$N(0, \sigX^2)$,
while $C$ has PDF
$N_c(0, \sigC^2)$,
where the variances are 
\be
\sigX^2 &=& \langle \vert X \vert^2 \rangle = 
	\Aave^2 \GiGjave_{BT}\, (BT)^{-1} 
	\left [
		B\,W_A m_A^2 
	      +  m_M^2 (1 + m_A^2) 
	\right ] \\
\sigC^2 &=&  \langle \vert C \vert^2 \rangle  = 
	(2\,BT)^{-1} 
	\left [ 
	   \Aave \left(\Niave\Gjave_{BT} + \Njave\Giave_{BT} \right) + 
		\Niave\Njave 
	\right ],
\ee
and we have used 
$\Gi \equiv \vert g_i \vert^2$, etc.
The forms of these variances are consistent with expressions given by
Rickett (1975). 
In general,  we can write the
variances of the real and imaginary parts of $X$ and $C$ as
$\sigma_{X_{r,i}}^2 = \half\sigma_X^2(1 \pm \rho_{G_{i,j}})$
and 
$\sigma_{C_{r,i}}^2 = \half\sigma_C^2(1 \pm \delta_{i,j})$.
When the DISS is perfectly correlated between the two sites,
the correlation coefficient $\rho_{G_{i,j}}$ (which is equal
to  $\gamG(\bvec, 0, 0, 0)$, 
c.f. Eq. \ref{eq:gamG-a}) is unity and $X$ is real.
As the DISS decorrelates between the sites, the 
$\sigma_{X_i} \to \sigma_{X_r}$.   
Only when the two sites are identical (e.g. for a single aperture
measurement of intensity) is $C$ real.   For all interferometers,
$C$ is complex with equal variances of the real and imaginary parts.

Deriving the variances involves assumptions about the correlation times
for the various signal terms and how they influence mean squares
of the time averages.    We have assumed that $n$ and $m$, the noise
processes, have correlation  times much smaller than that of the
amplitude modulation, $a$, which in turn has a much smaller correlation
time than the integration time used.  This hierarchy is consistent
with the fact that the noise correlation times are the reciprocal of the
bandwidth used.  For pulsars,  data are often obtained by using only
a small range of pulse phase but averaging over many pulse periods.
Pulsar pulses are broad band but decorrelate  on times about equal to
the spin period.  Our expression for $\sigma_X$ uses 
the correlation time $W_A$ for the amplitude
modulation.  This is effectively the width of the correlation function
$\rho_A$ defined in Eq.~\ref{eq:rhoAdef}.  As applied in Paper II, we would
take $W_A$ to be the width, $\Delta t \equiv P\Delta\phi_p$,
of the pulse window used in the analysis, where $P$ is the pulse phase
and $\Delta\phi_p$ is the window width in pulse phase units (cycles).
Then $T = N_p\Delta t $.   We have also included $m_M$, the modulation index
of pulsar noise fluctuations.  For amplitude modulated noise with Gaussian
statistics, $m_M \equiv 1$.   By retaining it, we can see what changes in the
statistics if we artifically turn off the noise fluctuations.    
$\sigC$ depends on $\Mave$ but not on $m_M$. 

\def\SNR{{\rm SNR}}
Note that the contribution of pulsar noise  to visibility fluctuations
relative to the contribution from additive noise is independent of
the averaging time.  Consider, for example, the ratio
$\sigma_X/\sigma_+$, where $\sigma_+$ is the value of $\sigma_C$ 
when there is no source.  We have $\sigma_X / \sigma_+ \propto \SNR_0$,
where $\SNR_0 \equiv \Aave / \sqrt{\Niave\Njave}$  is the ratio
of source strength to system temperature, when both are in the same
units (e.g. Janskys).  
Similarly, $\sigma_C/\sigma_+ - 1 \propto \sqrt{\SNR_0}$.  This result
is at odds with Gwinn \etal~(2000), who state that pulsar ``self noise'' 
can be ignored because it diminishes with averaging time.  
It does diminish but it cannot be ignored unless the signal to noise
ratio is small. 

The PDF for $\Gambar$ given $g_i(t)$ and  $g_j(t)$ 
may be calculated by appropriate
integration over the Gaussian PDFs for $X$ and $C$.   
We now consider some specific cases.

\subsection{DISS Perfectly Correlated Between Sites and 
Constant over $BT$}

For perfectly correlated DISS (between sites $i$ and $j$),
$\gigjave_{BT} \to \Giave_{BT} = \Gjave_{BT} \equiv \Gave_{BT}$ and
$\GiGjave_{BT} \to \G2ave_{BT}$;  thus $X$ becomes real.    
If, moreover, the DISS modulation is constant
over the averaging time $T$, then
$\Gave_{BT} \to G=$ constant and $\G2ave_{BT} \to G^2$.  The visibility
becomes 
\be
\Gambar = G\Aave + \Niave\delta_{ij} + X + C.
\ee
We assume a point source (and strong, saturated scintillations)
 so that, with $G$ constant over $T$, the
scintillation PDF is a one-sided exponential.
The PDFs of individual elements of $\Gambar$ are
\be
f_G(G) &=& e^{-G}U(G) \nonumber \\
f_X(X) &=& N(0, \sigX^2) \nonumber \\
f_C(C) &=& N_c(0, \sigC^2) \nonumber \\
\sigX^2 &=& G^2 \Aave^2 (BT)^{-1} 
	\left [ B\,W_A m_A^2 +  m_M^2 (1 + m_A^2) \right ] \nonumber \\ 
\sigC^2 &=& (2\,BT)^{-1} 
	\left [(\Niave \Njave) + 
		+  G \Aave (\Niave + \Njave ) 
	\right ].
\ee
%The DISS PDF, $f_G(G)$, is exponential in the saturated regime and for a
%point source, as we have assumed for this entire subsection.

\subsubsection{PDF of the Complex Visibility}

Let $\Gambar = \Gambar_r + i\Gambar_i$.    The PDF of $\Gambar$ is
\be
f_{\Gambar}(\Gambar) &=& 
	\int dG\, f_G(G) 
	\int dX f_X(X) f_C(\Gambar_r - G\Aave - X, \Gambar_i) \nonumber \\  
	&=&
	\int dG\, f_G(G) 
	\int dX (2\pi\sigma_X)^{-1/2} e^{-X^2/2\sigma_X^2}
	    (2\pi\sigma_C)^{-1}   
		e^{-\displaystyle\frac{1}{2\sigma_C^2}
		\left [ (\Gambar_r - G\Aave - X)^2 + \Gambar_i^2 \right ] }.
\ee
Performing the integral over $X$, we obtain
\be
f_{\Gambar}(\Gambar) &=& 
	(2\pi\sigma_C)^{-1} e^{-\Gambar_i^2/2\sigma_C^2}
	\int dG\, f_G(G)
	(\sigma_X^2 + \sigma_C^2)^{-1/2}
	e^{-\displaystyle\half (\Gambar_r - G\Aave)^2 /
                 2(\sigma_X^2 + \sigma_C^2)}.
\label{eq:complexPDF}
\ee
Note that for no signal ($\Aave \to 0$), we get
\be
f_{\Gambar}(\Gambar) &=&
        (2\pi\sigma_C^2)^{-1} e^{-\vert \Gambar\vert^2/2\sigma_C^2},
\ee
a circular Gaussian PDF.
Our expression in Eq.~\ref{eq:complexPDF}
disagrees with Eq. 11 of Gwinn \etal~(2000), which assigns equal
variances to the real and imaginary parts of $\Gambar$. The variances
are not equal, in general.  Also, there is an extra factor of $2\pi$ 
in their equation.

\subsubsection{PDF of the Visibility Magnitude}

The PDF of $\vert \Gambar \vert$ can be calculated as
\be
f_{\vert\Gambar\vert}(\vert\Gambar\vert) = 
	\vert\Gambar\vert \int_0^{2\pi}d\phi\,
	f_{\Gambar}(\vert\Gambar\vert\cos\phi, \vert\Gambar\vert\sin\phi).
\ee

Here we take a slightly different approach.
It is convenient to scale the magnitude of the  visibility 
by the rms of the complex term, $\sigC$.  
Using $\gamma \equiv \vert \Gambar \vert / \sigC$ and
$\inorm \equiv \Aave / \sigC$, the conditional PDF for
constant $G$ and $X=0$ is 
\be
\fgam(\gamma \vert \inorm, G) = 
	\gamma e^{-\half (\gamma^2 + G^2\inorm^2 )} 
	I_0(\gamma\inorm),
\label{eq:fgam1}
\ee
where $I_0$ is the modified Bessel function.  This result is the well
known Rice-Nakagami  PDF for a real phasor of length $\inorm$ combined with a
complex Gaussian phasor (e.g. Thomson, Moran \& Swenson 1991, Eq. 9.37).   
Integrating over the PDF for X, we have, for fixed $G$, 
\be
\fgam(\gamma | G) = 
	\int dX f_X(X) 
    	f_{\gamma}(\gamma \vert \inorm+{\displaystyle\frac{X}{G\sigC}}, G).
\label{eq:fgam2}
\ee
Then, integrating over the PDF for $G$, we have the PDF for $\gamma$
that takes into account all fluctuations, including DISS,
\be
\fgam(\gamma) = 
	\int dG f_G(G) \fgam(\gamma \vert  G).
\label{eq:fgam}
\ee

%\subsubsection{Infinite Signal-to-Noise Ratio}
%\label{sec:SNRlarge}

For S/N $\to \infty$, $\sigma_C \to 0$, 
\be
\Gambar &=& G\Aave + \Niave\delta_{ij} + X, 
\ee
and the PDF of $\vert \Gambar\vert$ for fixed $G$ becomes 
$N(G\Aave + \Niave \delta_{ij}, \sigX^2)$, with
the PDF for $\gamma$ given by Eq.~\ref{eq:fgam}.

\subsection{Perfectly Correlated DISS but $G\ne$ Constant over BT}

Specializing to the case of a weak source for which
$\sigX \ll \sigC$, we have (using $\Gbar \equiv  \Gave_{BT}$)
\be
\Gambar &\approx&  \Gbar\Aave + \Niave\delta_{ij}  + C \nonumber \\
\nonumber \\
f_{\Gbar}(\Gbar) &\approx& 
    \frac{(\Gbar\niss)^{\niss}} {\Gbar\Gamma(\niss)} e^{-\Gbar\niss} U(\Gbar), \nonumber
\ee
where $\Gamma(x)$ is the gamma function and  $U(x)$ is the unit step function. 
$\Gbar$ is distributed approximately 
 as $\chi^2_{2\niss}$, a chi-square
random variable  with $2\niss$ degrees of freedom, where 
$\niss \equiv \misssqinv$ and where $\misssq$ is given by Eq.~\ref{eq:ndof}. 
The true PDF of $\Gbar$ is obtained by solving the  
appropriate Fredholm equation for the eigenvalues that determine the
PDF, as described in the main text.

\begin{table}
\caption{Symbols and Acronyms Used}
\begin{center}

\begin{tabular}{| lll |}
\hline 
		&		&	\\
Symbols & Definition &  \\
		&		&	\\
\hline 
% top1
$\langle \cdots \rangle$ & Ensemble average & \\
$\langle \cdots \rangle_{BT} $ & Time average over time T of a 
           process with bandwidth B& \\
ACF		& Autocorrelation function & \\
CCF		& Crosscorrelation function & \\
DISS		& Diffractive Interstellar Scintillation & \\
DM		& Dispersion Measure &   \\
RISS		& Refractive Interstellar Scintillation & \\
$a(t)$		& Amplitude modulation (real) & \\
$A=a^2$		& Intensity modulation & \\
B, $\Delta\nu$ 	& Bandwidth & \\
$\bvec$		& Baseline	& \\
$b_e$		& 1/e scale of phase structure function & \\ 
$\beff$		& effective baseline & \\
$C_{\rm I}$	& Intensity CCF & \\
$\cnsq$		& Coefficient in electron-density wavenumber spectrum  &\\
d		& Earth-source distance &\\
$D = d-\ds$	& Screen-Earth distance &\\
$\ds$		& Source-screen distance &\\
$\dphi$		& Phase structure function & 	\\
$f_{\alpha}$	& Numerical factor in structure function &  \\		
$f_G(G)$	& PDF of scintillation modulation	 & \\
$f_{\gamma}(\gamma)$ & PDF of normalized visibility 	  &\\
$g$		& Wave propagator & \\
$G=\vert g \vert^2$	& Scintillation modulation of intensity or
				scintillation ``gain.''& \\
$E_{\Delta}$	& Narrowband electric field & \\
$\vareps$	& Complex baseband electric field & \\
$\vareps_s$	& Complex electric field at source & \\
I		& Intensity	&			\\
$\delta I$	& Intensity flucutation &		\\
$\Ibar$		& Time-average intensity & \\
${\cal L}$	& Likelihood function	& \\
$\ld$		& Characteristic spatial scale in diffraction pattern 	& \\
$m(t)$		& Complex gaussian noise with unit mean square& \\
$M \equiv \vert m\vert^2$ & Squared magnitude of $m$  & \\
$m_A$		& Modulation index of A	& \\
\hline 
\end{tabular}
\end{center}
\end{table}

\addtocounter{table}{-1}
\begin{table}
\caption{Symbols and Acronyms Used (continued)}
\begin{center}
\begin{tabular}{| lll |}
\hline 
		&		&	\\
Symbols & Definition & \\
		&		&	\\
\hline 
% top2
$\miss$		& Scintillation modulation index (= rms / mean) & \\
$m_M$		& Modulation index of M & \\
$\ndof$		& Number of degrees of freedom in scintillations & \\
$\niss$		& Number of independent scintillation features averaged & \\
$n(t)$		& Additive complex Gaussian noise & \\
$n_e$		& Free electron density		&  	  \\ 
$R_{\Ibar}$	& ACF of averaged intensity & \\
$R_{\Gambar}$	& ACF of averaged visibility & \\
$R_{4\vareps}$	& Fourth moment of complex field & \\
$\rsvec$	& Two dimensional vector at source & \\
$r_0$		& Fried scale 		& \\
$s$		& Location along line of sight, $s=0$ at source & \\
$T$		& Integration or averaging time & \\
$\tau_i$	& Time lag used in calculating interferometer visibilities & \\
$\veffvec$	& Vector effective velocity & 	\\
$\vobs$		& Observer's velocity	&		\\
$\vp$		& Pulsar velocity	&		\\
$\vism$		& Velocity of scattering medium 	& \\
$\viss$		& Velocity of ISS diffraction pattern	& \\
$\alpha$	& Exponent in phase structure function		& \\
$\Gamma_{\rm I}$	& Intensity autocovariance function &	\\
$\Gamma_{\vareps}$	& Visibility function & \\
$\Gambar$	& Time average of visibility function & \\
$\gamg$		& Second moment of propagator $g$ & \\
$\gamG$		& autocovariance of scintillation `gain' $G$ & \\
$\Delta(\tau)$	& Normalized ACF of noise & \\
$\dnud$		& Diffraction or scintillation bandwidth 	& \\
$\dtd$		& Diffraction or scintillation time scale & \\
$\phi$		& Phase perturbation from refractive index perturbations &  \\
$\psi_n$	& Eigenfunction in Karhunen-Lo\`eve problem & \\		
$\rho_A$	& Autocorrelation function of $A(t)$ & \\
$\biso$		& Isoplanatic scale of diffraction pattern at observer's
			location & \\
$\drsiso$	& Isoplanatic scale at source's location & \\
$\sigma_r$	& Length scale in Gaussian brightness distribution & \\ 
$\thiso$	& Isoplanatic angle  & \\
$\ths$		& Source angular size & \\
\hline 
\end{tabular}
\end{center}
\end{table}


\end{document}